\documentclass[camera]{jpaper}
\usepackage{cite}

\usepackage{soul}

\usepackage{amsmath}
\usepackage{graphicx}
\usepackage[linesnumbered,ruled]{algorithm2e}
\usepackage{algpseudocode}
\usepackage{titlesec}
\algrenewcommand\algorithmiccomment[2][\normalsize]{{#1\hfill\(\triangleright\) #2}}
\usepackage{geometry} \titlespacing*{\section}{0pt}{3pt plus 1pt minus 1pt}{1pt}
\titlespacing*{\subsection}{0pt}{3pt plus 1pt minus 1pt}{1pt}
\titlespacing*{\subsubsection}{0pt}{1pt}{0pt}

\usepackage{array}
\makeatletter
\let\MYcaption\@makecaption
\makeatother

\usepackage[font=footnotesize]{subcaption}

\makeatletter
\let\@makecaption\MYcaption
\makeatother

\usepackage{fixltx2e}

\usepackage{dblfloatfix}
\usepackage[nolessnomore, italic]{mathastext}
\usepackage[T1]{fontenc}
\usepackage[usenames,dvipsnames,svgnames,table]{xcolor}
\usepackage{multirow}
\usepackage{hhline}
\usepackage[normalem]{ulem}
\usepackage{setspace}
\usepackage{indentfirst}
\usepackage{footmisc}

\usepackage{pifont}
\usepackage{footnote}
\makesavenoteenv{tabular}
\makesavenoteenv{table}

\newif\ifcameraready
\camerareadytrue

\ifcameraready

\else

\fi
\newcommand{\affilETH}[0]{\textsuperscript{\S}}

\newcommand{\affilShanghaiTech}[0]{\textsuperscript{$\dagger$}}

\newcommand*\circled[1]{\tikz[baseline=(char.base)]{\node[shape=circle,fill,inner sep=.8pt] (char) {\textcolor{white}{#1}};}}

\newcommand*\hcircled[1]{\tikz[baseline=(char.base)]{\node[shape=circle, draw=black, fill=white, inner sep=.8pt] (char) {\textcolor{black}{#1}};}}

\newcommand{\blbar}[0]{$\overline{\text{bitline}}$}
\newcommand{\bl}[0]{$\text{bitline}$}

\definecolor{amber}{rgb}{1.0, 0.49, 0.0}
\definecolor{darkgreen}{rgb}{0.0, 0.2, 0.13}
\definecolor{darkbyzantium}{rgb}{0.36, 0.22, 0.33}
\definecolor{darkseagreen}{rgb}{0.56, 0.74, 0.56}
\definecolor{darkspringgreen}{rgb}{0.09, 0.45, 0.27}
\definecolor{dollarbill}{rgb}{0.52, 0.73, 0.4}
\definecolor{midnightblue}{rgb}{0.0, 0.26, 0.37}
\definecolor{porsche}{rgb}{0.918, 0.647, 0.337}

\newcommand{\mechanism}[0]{CLR-DRAM}

\newif\ifsubmission
\usepackage{dblfloatfix}
\usepackage{graphicx}
\usepackage{lipsum}
\usepackage{upgreek}
\usepackage{tikz}
\usepackage{xcolor}
\usepackage{soul}
\usepackage{ifthen}

\submissionfalse

\ifsubmission

\newcommand{\ts}[1]{}
\newcommand{\taha}[1]{}

\newcommand{\lois}[1]{{\color{blue}#1}}
\newcommand{\hh}[1]{{\color{blue}#1}}

\newcommand{\agycomment}[1]{}

\else

\newcommand\ts[1]{\noindent{\color{blue}{\bf\textbf{}}~{\it Taha: #1}}~} 
\newcommand{\taha}[1]{{\color{blue}#1}}

\newcommand{\jsc}[2]{\ifthenelse{\equal{#1}{10}}{{\color{purple}#2}}{{\color{black}#2}}}
    
\newcommand{\lois}[1]{{\color{blue}#1}}
\newcommand{\hh}[1]{#1}

\newcommand{\mpg}[1]{{\color{dollarbill}#1}}
\newcommand{\agycomment}[1]{\noindent{\color{RedOrange}{\textbf{@giray:~{#1}}}}}

\fi

\begin{document}
\bstctlcite{IEEEexample:BSTcontrol} 


\title{\mechanism{}: A Low-Cost DRAM Architecture\\Enabling Dynamic Capacity-Latency Trade-Off}


%


\author{
{Haocong Luo\affilETH\affilShanghaiTech}\qquad%
{Taha Shahroodi\affilETH}\qquad%
{Hasan Hassan\affilETH}\qquad%
{Minesh Patel\affilETH}\qquad \\%
{A. Giray Ya\u{g}l{\i}k\c{c}{\i}\affilETH}\qquad
{Lois Orosa\affilETH}\qquad%
{Jisung Park\affilETH}\qquad%
{Onur Mutlu\affilETH}\qquad\vspace{-3mm}\\\\
{\vspace{-3mm}\affilETH \emph{ETH Z{\"u}rich} \qquad \affilShanghaiTech \emph{ShanghaiTech University}}%
}


%


\maketitle
\thispagestyle{plain} 
\pagestyle{plain}

\setstretch{0.934}
\renewcommand{\footnotelayout}{\setstretch{0.9}}








%


\begin{abstract}

\par  DRAM is the prevalent main memory technology, but its long access latency can limit the performance of many workloads.
Although prior works provide DRAM designs that reduce DRAM access latency, their reduced storage capacities hinder the performance of workloads that need large memory capacity.  Because the capacity-latency trade-off is fixed at design time, previous works cannot achieve maximum performance under very different and dynamic workload demands.


\par This paper proposes Capacity-Latency-Reconfigurable DRAM (\mechanism{}), a new DRAM architecture that enables dynamic capacity-latency trade-off at low cost. 
\mechanism{} allows dynamic reconfiguration of any DRAM row to switch between two operating modes: 1) max-capacity mode, where every DRAM cell operates individually to achieve approximately the same storage density as a density-optimized commodity DRAM chip 
and 2) high-performance mode, where two adjacent DRAM cells in a DRAM row and their sense amplifiers are coupled to operate as a single  low-latency logical cell driven by a single logical sense amplifier.

\par We implement \mechanism{} by adding isolation transistors in each DRAM subarray.
Our evaluations show that \mechanism{} can improve system performance and DRAM energy consumption by 18.6\% and 29.7\% on average with four-core multiprogrammed workloads. We believe that \mechanism{} opens new research directions for a system to adapt to the diverse and dynamically changing memory capacity and access latency demands of workloads.

\end{abstract}
\section{Introduction}

\par 
DRAM is the prevalent technology for architecting main memory in modern computing systems. During the past two decades, the storage capacity of a commodity DRAM chip has increased by more than two orders of magnitude (e.g., from 128\,Mb~\cite{Samsung-ddr-128Mb} to 16\,Gb~\cite{Samsung-ddr4-16Gb}) to meet the increasing main memory capacity demands of modern applications. 
In contrast, DRAM access latency\footnote{Commonly measured by \emph{\textbf{tRC}} (Row Cycle time)~\cite{Samsung-ddr-128Mb, Samsung-ddr4-16Gb}.} has reduced by only 16.7\% over the same period~\cite{lee2013tiered, FLYDRAM, chang2017understanding, borkar-cacm2011, hassan2016chargecache, lee2016reducing, lee2016thesis, chang2017understandingPHD}.
Modern processors spend hundreds of clock cycles to access data stored in DRAM, which leads to a significant system-level performance bottleneck~\cite{Son2013CHARM, mutlu2013memory, Hestness2014ACA, wilkes2001memory, wulf1995hitting, mutlu2007stall, mutlu03runahead, kanev2015profiling, Boroumand2018Google, mutlu2014superfri, koppula2019eden, bera2019dspatch, kanellopoulos2019smash, liu2019binary, ghose2019processing}.

\par
The long access latency is not intrinsic to the DRAM technology itself; there is a fundamental trade-off between access latency and storage density in DRAM. For example, vendors optimize commodity DRAM devices for high storage density at the cost of long access latency. In contrast, vendors also offer special-purpose DRAM devices optimized for low latency but with significantly lower storage density compared to commodity DRAM~\cite{rldram, sato1998fast}. 
Recent works attempt to provide the best of both density- and latency-optimized DRAM by developing heterogeneous DRAM architectures~\cite{choi2015multiple, lee2013tiered, LISA, Son2013CHARM} that provide low-latency access in a \emph{small} and \emph{fixed} region within a DRAM chip. Unfortunately, all of these works make the capacity-latency trade-off \emph{statically} at design time: the amount of density-optimized or latency-optimized DRAM is fixed once the chip is fabricated.
\par
We observe that existing DRAM architectures that make the DRAM capacity-latency trade-off decision \emph{statically} at design-time~\cite{lldram,hpdaram,rldram, sato1998fast, lee2013tiered,Kim2006TwinCell,twincell, takemura20060, choi2015multiple} \emph{cannot} adapt to changes in a system's main memory capacity and latency demands, which vary over time with workload behavior. Therefore, existing systems miss opportunities to improve performance. For example, commodity density-optimized DRAM is unable to reduce DRAM access latency when applications exhibit low memory capacity demand. Similarly, latency-optimized DRAM cannot extend its memory capacity when memory capacity is insufficient and frequent page faults lead to significant performance degradation. Existing heterogeneous DRAM architectures~\cite{choi2015multiple, lee2013tiered, LISA, Son2013CHARM} take a step towards providing the best of both density- and latency-optimized DRAM. However, they leave significant potential for improving DRAM access latency untapped because they employ a \emph{fixed-size} and small low-latency region that does not always provide the best possible operating point within the DRAM capacity-latency trade-off spectrum for all workloads.

\begin{figure}[h]
    \centering
    \hspace*{-.1in}
    \includegraphics[width=0.5\textwidth]{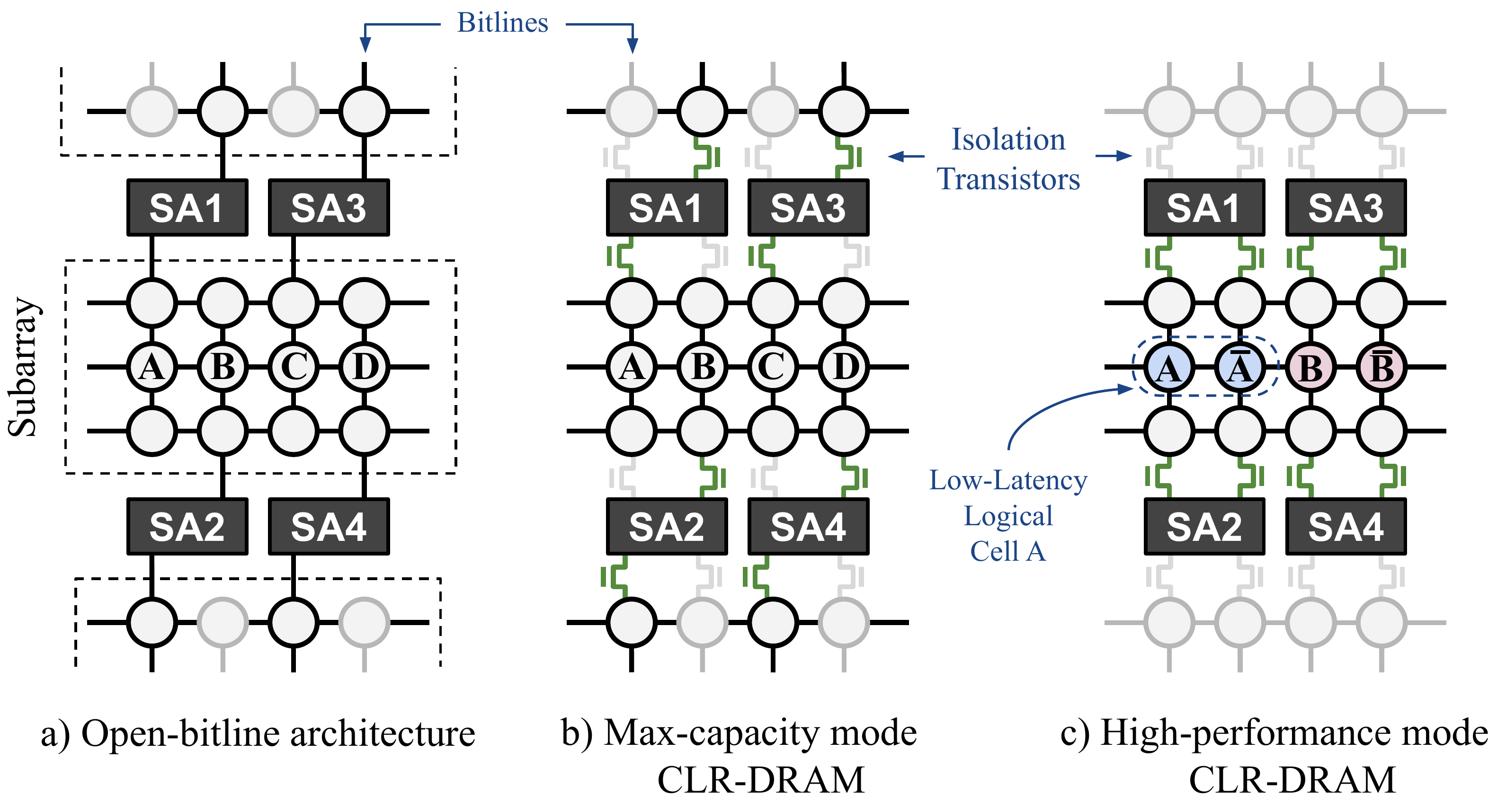}

    \caption{Comparison of a) the conventional open-bitline architecture and our \mechanism{} architecture operating in b) max-capacity mode and c) high-performance mode. Isolation transistors highlighted in \textbf{green} are enabled, faded grey disabled.}

    \label{fig:OB_Arch_Intro}
    \vspace*{-.09in}

\end{figure}

\textbf{Our goal} is to design a low-cost DRAM architecture that can be dynamically configured to have high capacity or low latency at a fine granularity (i.e., at the granularity of a row). 
To this end, we propose \mechanism{} (Capacity-Latency Reconfigurable DRAM), which extends the conventional density-optimized open-bitline DRAM architecture (Figure~\ref{fig:OB_Arch_Intro}a) by adding isolation transistors along the bitlines in each DRAM subarray (Figures~\ref{fig:OB_Arch_Intro}b and \ref{fig:OB_Arch_Intro}c). \mechanism{} is able to dynamically control how DRAM cells are connected with their corresponding sense amplifiers at the granularity of a DRAM row. \textbf{The key idea} of \mechanism{} is to enable the ability to \emph{dynamically} reconfigure \emph{any single DRAM row} to operate in either max-capacity mode (Figure~\ref{fig:OB_Arch_Intro}b) or high-performance mode (Figure~\ref{fig:OB_Arch_Intro}c).


\par The max-capacity mode (Figure~\ref{fig:OB_Arch_Intro}b) operates exactly the same as in the conventional open-bitline architecture (Figure~\ref{fig:OB_Arch_Intro}a) by enabling a subset of the newly-added isolation transistors (shown in green) to mimic the conventional design's interconnect. Doing so achieves the same storage capacity as an unmodified capacity-optimized DRAM chip. The high-performance mode (Figure~\ref{fig:OB_Arch_Intro}c) couples every two \emph{adjacent} physical DRAM cells in the same row along with their two sense amplifiers to operate \emph{together as a single low-latency logical cell} and a single, but stronger, logical sense amplifier by enabling \emph{all} of the newly-added isolation transistors. 

\par The two coupled physical cells (e.g., $\textbf{A}$ and $\overline{\textbf{A}}$ in Figure~\ref{fig:OB_Arch_Intro}c) store the \emph{same} bit of data encoded using \emph{opposite} charge levels. The coupled cells connect to \emph{both} ports (\emph{bitline} and \emph{\blbar}) of a sense amplifier, taking advantage of the \emph{differential} operation principle of the sense amplifier to begin its sensing process \emph{much earlier} than in the max-capacity mode (and the conventional open-bitline architecture). 




\par
The two coupled sense amplifiers (e.g., \textbf{SA1} and \textbf{SA2} in Figure~\ref{fig:OB_Arch_Intro}c) accelerate a DRAM access in three ways. First, they drive the logical cell formed by the two coupled cells simultaneously from \emph{both} ends of the bitlines, reducing the latency of a DRAM row activation.
Second, coupling the two sense amplifiers makes their precharge units operate together to reduce the latency of precharge operations, which prepare the DRAM bank for activating a new row.
Third, since refreshing a DRAM row is analogous to performing an activation and precharge on the row~\cite{venkatesan2006retention, liu2012raidr, nair2013archshield, qureshi-dsn2015, chang2014improving}, a row operating in high-performance mode is refreshed with low refresh latency, which reduces the performance and energy-consumption overheads of DRAM refresh.

\par
We enable two optimizations for DRAM rows operating in high-performance mode, where every two coupled cells store opposite levels of charge. First, we leverage the observation that a cell storing a low level of charge can safely terminate the trailing charge-restoration phase of a DRAM row activation \emph{earlier} than a cell storing a high level of charge. Doing so reduces the DRAM row charge-restoration latency (\textbf{tRAS}) in high-performance mode even further, thereby improving performance. Second, we exploit the high effective capacitance of the two coupled cells storing opposite charge levels to extend the interval between consecutive refresh operations for a row operating in high-performance mode.

\par Our detailed circuit-level simulations show that, with all our optimizations enabled, \mechanism{} significantly reduces DRAM access latency with low overheads.
\mechanism{} allows reducing the activation latency (\textbf{tRCD}) by 60.1\%, restoration latency (\textbf{tRAS}) by 64.2\%, write recovery latency (\textbf{tWR}) by 35.2\%, and precharge latency (\textbf{tRP}) by 46.4\% while incurring a modest DRAM chip area overhead of at most 3.2\%.
\par
Our evaluations using 41 single-core and 90 four-core multiprogrammed workloads show that \mechanism{} significantly improves performance and energy efficiency.
For multi-core workloads, if we map only 25\% of the most-accessed memory pages to high-performance rows, \mechanism{} provides an average performance improvement of 11.9\% while reducing average DRAM energy and power consumption by 21.7\%  and 8.9\%, respectively.
If we map \emph{all} memory pages to high-performance rows, the performance gain, DRAM energy savings, and DRAM power reduction  further increase to 18.6\% (27.5\% for memory-intensive workloads), 29.7\% and 12.8\%, respectively.

\par
We conduct a sensitivity analysis of extending the refresh interval to show that we can safely extend the refresh interval of DRAM rows operating in high-performance mode by up to 3$\times$. If all DRAM rows are configured to operate in high-performance mode, \mechanism{} reduces the energy consumption of DRAM refresh commands by 87.1\% while still providing an average performance improvement of 17.8\% in multi-core evaluations.

\par This paper makes the following key contributions:
\begin{itemize}[leftmargin=*]
    \item We propose \mechanism{}, the first DRAM architecture that enables a \emph{dynamic} trade-off between storage capacity and low-latency operation in DRAM at the fine granularity of {\emph{a single DRAM row} with \emph{low} hardware cost. 
    }
    
    \item We show that the conventional open-bitline architecture, originally designed for high DRAM storage capacity, can be exploited to enable the coupled operation of two DRAM cells and two sense amplifiers. Such coupling significantly reduces \emph{four} major DRAM access latencies (\textbf{tRCD}, \textbf{tRAS}, \textbf{tRP} and \textbf{tWR}) as well as DRAM refresh overhead. 
    \item We evaluate \mechanism{}'s system-level performance, energy, and power consumption to show that it 
    significantly improves all three metrics across both single-core and multi-programmed workloads.
    
\end{itemize}



\section{DRAM Background}
\label{sec:background}


We provide background on DRAM organization and operation for understanding the details of our proposal. We refer the reader to prior studies~\cite{Hassan2019CROW, ghose2019demystifying, ghose2018your, kim2015ramulator, seshadri2019dram, kim-isca2012, zhang2014half, hassan2016chargecache, lee2013tiered, seshadri2017ambit, chang2017understanding, FLYDRAM, chang2014improving, LISA, AL-DRAM, lee2016reducing, lee2016thesis, koppula2019eden, lee2015decoupled, liu2013experimental, liu2012raidr, seshadri2013rowclone, seshadri2015gather, ipek2008self, lee2016simultaneous, Dennard68field} for a more detailed description of DRAM architecture.

\subsection{DRAM Organization}
\label{sec:organization}
\par
The DRAM-based main memory of a modern computer system is organized hierarchically. A CPU accesses DRAM through a memory controller. The DRAM can be organized in multiple DRAM \emph{channels}, and the memory controller communicates with each channel using a separate \emph{channel bus} that allows operating the channel independently from the others. A channel contains multiple DRAM \emph{ranks}, where each rank shares the channel bus in a time-multiplexed manner. A rank consists of a set of DRAM \emph{chips} that operate in lock-step. A DRAM chip is partitioned into multiple DRAM \emph{banks}, which process different DRAM requests simultaneously. 

\par
\begin{figure}[!ht]
\centering
\includegraphics[width=0.9\linewidth]{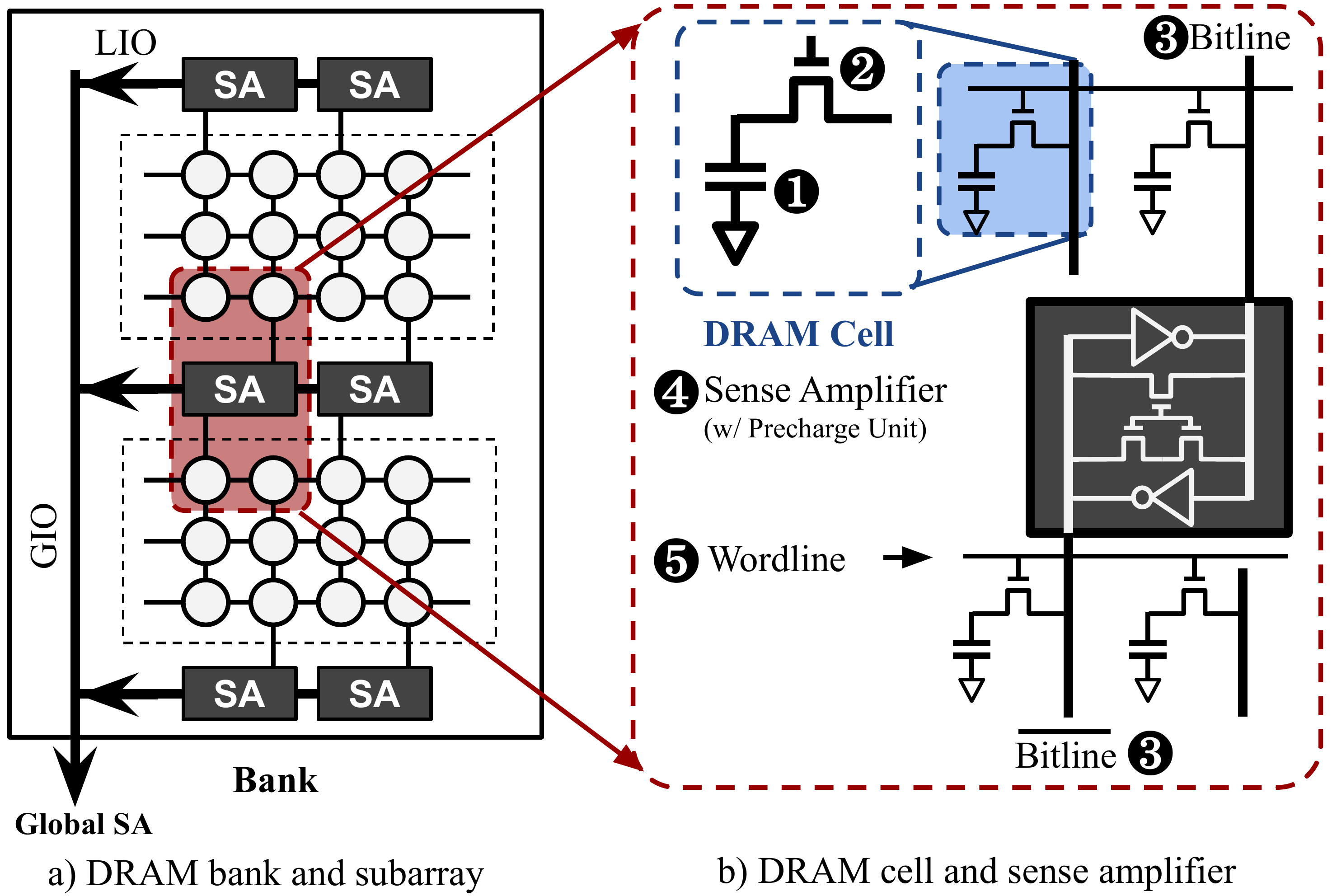}
\caption{DRAM organization.}
\label{fig:DRAM_ORG}
\end{figure}

 A DRAM bank consists of multiple (typically between 128 to 512) \emph{subarrays} 
as illustrated in Figure~\ref{fig:DRAM_ORG}a. A subarray comprises a two-dimensional DRAM cell array, and each column of cells in this array is connected to a \emph{sense amplifier (SA)}. SAs in each subarray are connected to Global IO (GIO) wires via Local IO (LIO) wires. GIO wires transfer a column of data between SAs of the subarrays and the bank-level Global SA.

In Figure~\ref{fig:DRAM_ORG}b, we take a closer look into a DRAM cell and how it is  connected to a sense amplifier. A DRAM \emph{cell} contains a capacitor \circled{1} that stores one bit of information in the form of electrical charge (e.g., charged/discharged capacitor representing logical `1'/`0' or vice versa) and  an access transistor \circled{2} that connects the capacitor to  a \emph{bitline} \circled{3} (or \emph{\blbar}). Cells in a column share a bitline, which connects them to a  sense amplifier (SA) \circled{4} at the end of the bitline. Similarly, cells in a row share  a \emph{wordline} \circled{5} that drives the gate of each cell's access transistor. To access data stored in a DRAM cell, the wordline is asserted to enable the charge of a cell capacitor to be shared with the corresponding SA via the bitline.

A sense amplifier (SA) is responsible for amplifying the small amount of electrical charge stored in a DRAM cell to an accessible level. A DRAM SA operates based on the difference in the voltage levels of the pair of bitlines (\bl~and \blbar) that are connected to the SA. Once a DRAM cell shares its small amount of charge with one of the two bitlines, the SA amplifies this small perturbation across the bitline pair to an accessible voltage level. An SA also includes a \emph{precharge unit that resets the SA and the bitline pair to prepare for access to another DRAM row. We explain the details of a DRAM access in Section~\ref{subsec:dram_operation}.} Throughout this paper, we use the term ``sense amplifier (SA)'' to refer to \emph{both} the SA and the precharge unit.

\noindent \textbf{Open-bitline Architecture.}
Due to the large size of an SA, the relative arrangement of SAs and DRAM cells significantly impacts the storage density of a DRAM chip. A common  density-optimized bank design is known as the \emph{open-bitline Architecture}~\cite{LISA,keeth2007dram}, which places SAs on both sides (i.e., top and bottom) of a subarray as we show in Figure~\ref{fig:DRAM_ORG}a. In this architecture, neighboring DRAM cell are connected to SAs placed on different sides of the subarray through a pair of adjacent bitlines. For example, in Figure~\ref{fig:DRAM_ORG}a, DRAM cell \textbf{A}  is connected to \textbf{SA1} at the top of the subarray while its neighboring cell \textbf{B} is connected to \textbf{SA2} at the bottom of the subarray.

\subsection{DRAM Operation}
\label{subsec:dram_operation}

A DRAM access requires three main steps: \emph{row activation}, \emph{column access}, and \emph{precharge}.

\begin{figure}[ht]
\centering
\includegraphics[width=0.47\textwidth]{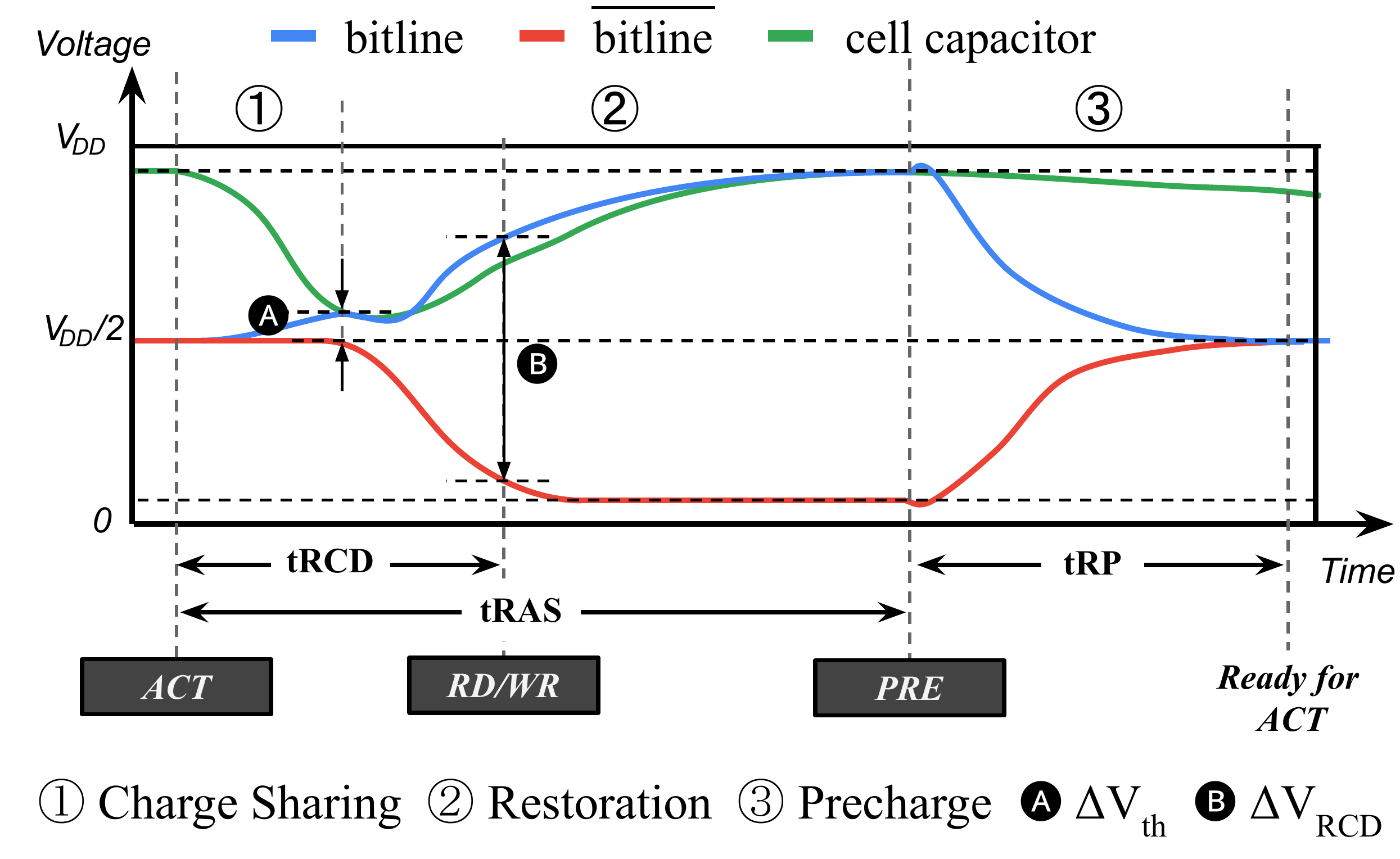}

\caption{Bitline and cell capacitor voltages during a DRAM access.}
\label{fig:DRAM_OP}
\end{figure}

\noindent \textbf{Row Activation.} 
To perform read/write operations on data stored in DRAM cells, stored data must first be loaded into the SAs by activating the row to be accessed. As we show in Figure~\ref{fig:DRAM_OP}, the row activation process is composed of two phases:  \emph{charge sharing} \lois{\hcircled{1}} and  \emph{charge restoration} \lois{\hcircled{2}}. 
Prior to row activation, the bitlines are 
in a \emph{precharged} state at the reference voltage level (e.g., $V_{DD}/2$). To activate a row, the memory controller issues a row activation command \textbf{ACT} to the DRAM.
The first phase of row activation, charge sharing \hcircled{1}, starts by asserting the wordline of a row, leading to charge sharing between the bitlines and the capacitors of all the cells along the activated row. Charge sharing slightly perturbs the bitlines and causes a small difference in the bitline voltage~\circled{A}. During the charge restoration phase \hcircled{2}, each SA amplifies the small difference in bitline voltage to a level suitable for read/write access. Charge restoration also replenishes the charge in the cell capacitor since, during this phase, the wordline remains asserted, and thus the cell capacitor remains connected to the bitline. The DRAM standard specifies the time interval from the beginning of row activation until the two phases complete as \textbf{tRAS}~\cite{jedec-ddr4, AL-DRAM, kim-isca2012, lee2013tiered}.

\noindent \textbf{Column Access.} The memory controller can issue a column access command (i.e., read, write) during the  charge-restoration phase \lois{\hcircled{2}} once the voltage difference across the bitlines reaches the ``ready-to-access'' voltage level \circled{B}. The DRAM standard specifies the time interval between the start of row activation and when the bitline reaches the ``ready-to-access'' voltage level \circled{B} as \textbf{tRCD}~\cite{jedec-ddr4, AL-DRAM, kim-isca2012, lee2013tiered}.

\noindent \textbf{Precharge.} To activate a new DRAM row, the memory controller first issues a precharge command (\textbf{PRE}) to  precharge the activated row \hcircled{3} and prepare the bank for a subsequent row activation. The precharge operation de-asserts the wordline to disconnect the activated row of cells from the bitlines and resets the bitlines back to the reference voltage (e.g., $V_{DD}/2$). The DRAM standard specifies the time interval from issuing the precharge command until the precharge completes as \textbf{tRP}~\cite{jedec-ddr4}.


\noindent \textbf{Refresh.} A DRAM cell capacitor's charge leaks over time. To ensure that DRAM retains its data indefinitely, the charge in all cell capacitors in a DRAM chip must be periodically replenished. The memory controller ensures data retention by periodically issuing DRAM refresh commands. The DRAM standard specifies the time interval between two consecutive refresh commands that refresh the same row as the refresh window \textbf{tREFW}. A single refresh command refreshes multiple rows, and the number of refreshed rows depends on the total number of rows in a bank. The DRAM standard further specifies the latency of a refresh command (i.e., time required to complete a single refresh operation) as \textbf{tRFC}~\cite{jedec-ddr4, liu2012raidr, patel2017reach}.

\subsection{Sense Amplifier}
\label{SA}
A DRAM sense amplifier (SA) is connected to a \emph{pair} of bitlines, $bitline$ and $\overline{bitline}$, through its two \emph{complementary} ports (also denoted as $bitline$ and $\overline{bitline}$). During an access to the DRAM cell, the bitline pair develops from a reference voltage (e.g., $V_{DD}/2$) into complementary voltage levels (e.g., to $\{V_{bitline} = V_{DD}$, $ V_{\overline{bitline}} = 0\}$).

\par Prior to a row activation, the precharge unit in the SA sets the voltage of the bitlines to a reference value  and the SA is disabled. Then, the charge sharing phase (\hcircled{1} in Figure~\ref{fig:DRAM_OP}) introduces a small voltage difference ($\Delta V$) across the bitline pair (\circled{A} in Figure~\ref{fig:DRAM_OP}) by letting the cell share its charge with a bitline. In Figure~\ref{fig:DRAM_OP}, the capacitor of the cell connected to $bitline$ is initially fully charged. Charge sharing \hcircled{1} then perturbs $bitline$, making its voltage $V_{bitline}$ slightly higher than $V_{\overline{bitline}}$. After a certain amount of time spent in charge sharing \mpg{\hcircled{1}} to ensure that $\Delta V$ reaches a threshold ($\Delta V = \Delta V_{th}$  \circled{A}), internal control circuitry enables the SAs. Then, the SA amplifies $\Delta V$ so that $V_{bitline}$ reaches $V_{DD}$ and $V_{\overline{bitline}}$ reaches $0$. 



\section{\mechanism{} Architecture}
\par
We propose \mechanism{}, a new DRAM architecture that enhances the conventional open-bitline DRAM architecture with isolation transistors to support dynamic reconfiguration of the connections between DRAM cells and SAs (Section~\ref{Design}).

\par \mechanism{} enables
\emph{dynamic} reconfiguration of \emph{any given DRAM row} to operate in either max-capacity mode or high-performance mode (Section~\ref{twomodes}) to better meet varying capacity and performance requirements.
\mechanism{} requires only two control signals (and their complements) per-bank for controlling the isolation transistors across \emph{all} subarrays of the bank to support row-level reconfigurability between the two modes (Section~\ref{Control}). 
 
\par High-performance mode improves DRAM cell access performance in three ways. First, it reduces three major cell access latency timings: \textbf{tRCD}, \textbf{tRAS}, and \textbf{tRP} (discussed in Section \ref{Operation}). Second, it enables early termination of charge sharing to \emph{further} reduce \textbf{tRAS} and \textbf{tWR} (Section \ref{earlyprecharge}). Third, it reduces the latency of a refresh operation (\textbf{tRFC}) and increases DRAM cells' retention times to mitigate performance and energy costs associated with DRAM refresh (Section \ref{reducerefresh}).

\subsection{Bitline Mode Select Transistors}\label{Design}

%




 \par
\begin{figure}[!h]
\vspace{1pt}
\centering
\includegraphics[width=0.75\linewidth]{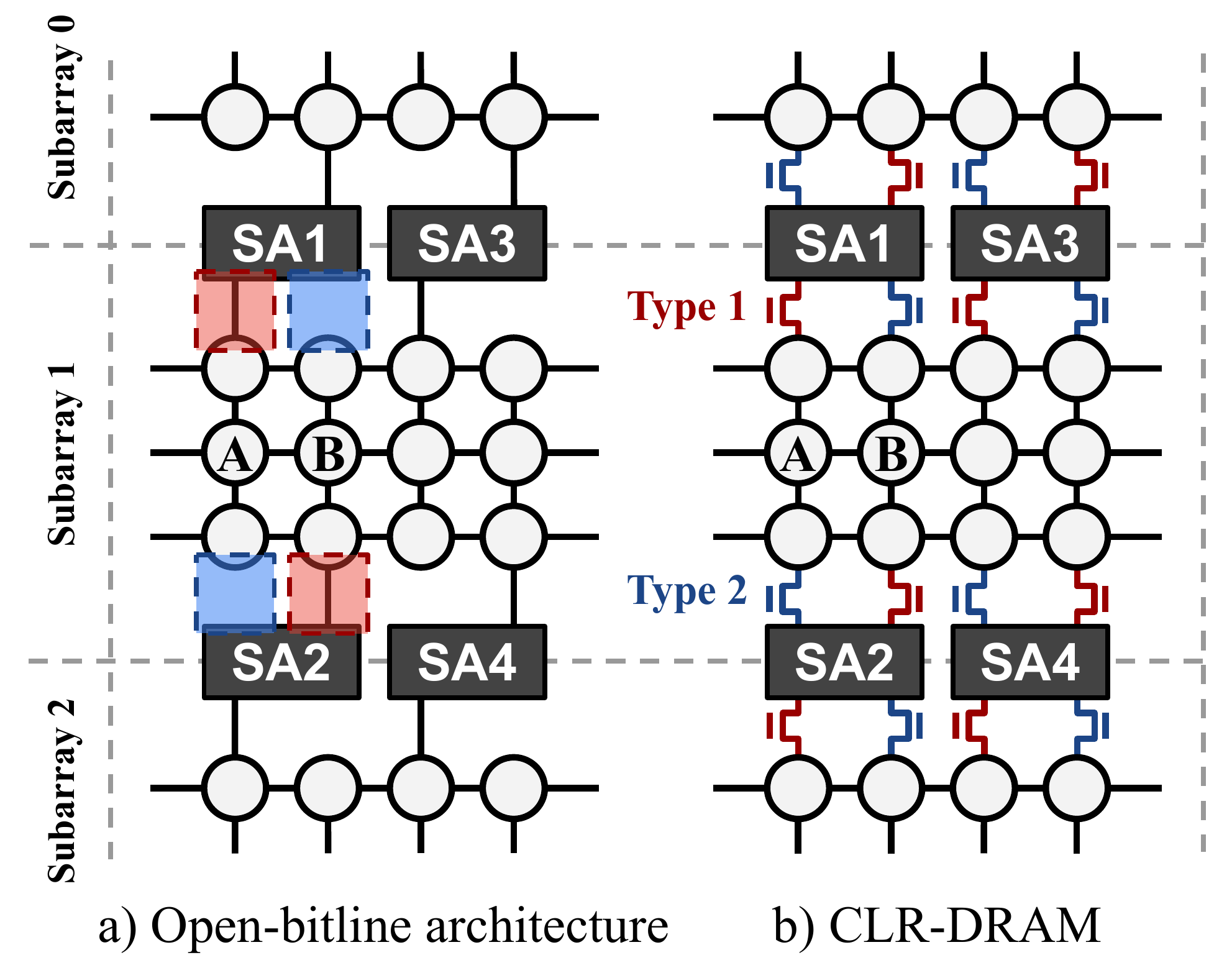}
\caption{Baseline open-bitline architecture vs. \mechanism{}.}
\label{fig:TCTS_Arch}
\end{figure}

 \mechanism{} requires small modifications to the state-of-the-art density-optimized open-bitline architecture (Figure~\ref{fig:TCTS_Arch}a). We make two observations about the open-bitline architecture that enable implementing \mechanism{} at low cost. First, two adjacent DRAM cells in a row (e.g., cells \textbf{A} and \textbf{B} in Figure~\ref{fig:TCTS_Arch}a) are connected to two \emph{different} SAs (cell \textbf{A} to \textbf{SA1} and cell \textbf{B} to \textbf{SA2}) at different sides of the subarray (top and bottom). Second, one end of the bitline is connected to the SA of one side of the subarray (marked in red), while the other end of the bitline is close to, but not connected to, the SA on the other side of the subarray (marked in blue).

\par Leveraging these observations, we implement \emph{\mechanism{}} to enable dynamic reconfiguration of the connections between DRAM cells and SAs (Figure~\ref{fig:TCTS_Arch}b) by adding an isolation transistor, called \emph{bitline mode select transistor}, to \emph{each end of every bitline}. 
A \textbf{Type 1} transistor (colored red) replaces a previous \emph{connection} between a bitline and a sense amplifier (e.g., between cell \textbf{A} and \textbf{SA1}), and a  \textbf{Type 2} transistor (colored blue) connects a previously \emph{unconnected} end of a bitline to a sense amplifier (e.g., bitline of cell \textbf{B} to \textbf{SA1}). 


\subsection{Max-Capacity and High-Performance Modes}\label{twomodes}


\par
\begin{figure}[h]
\centering
\includegraphics[width=0.43\textwidth]{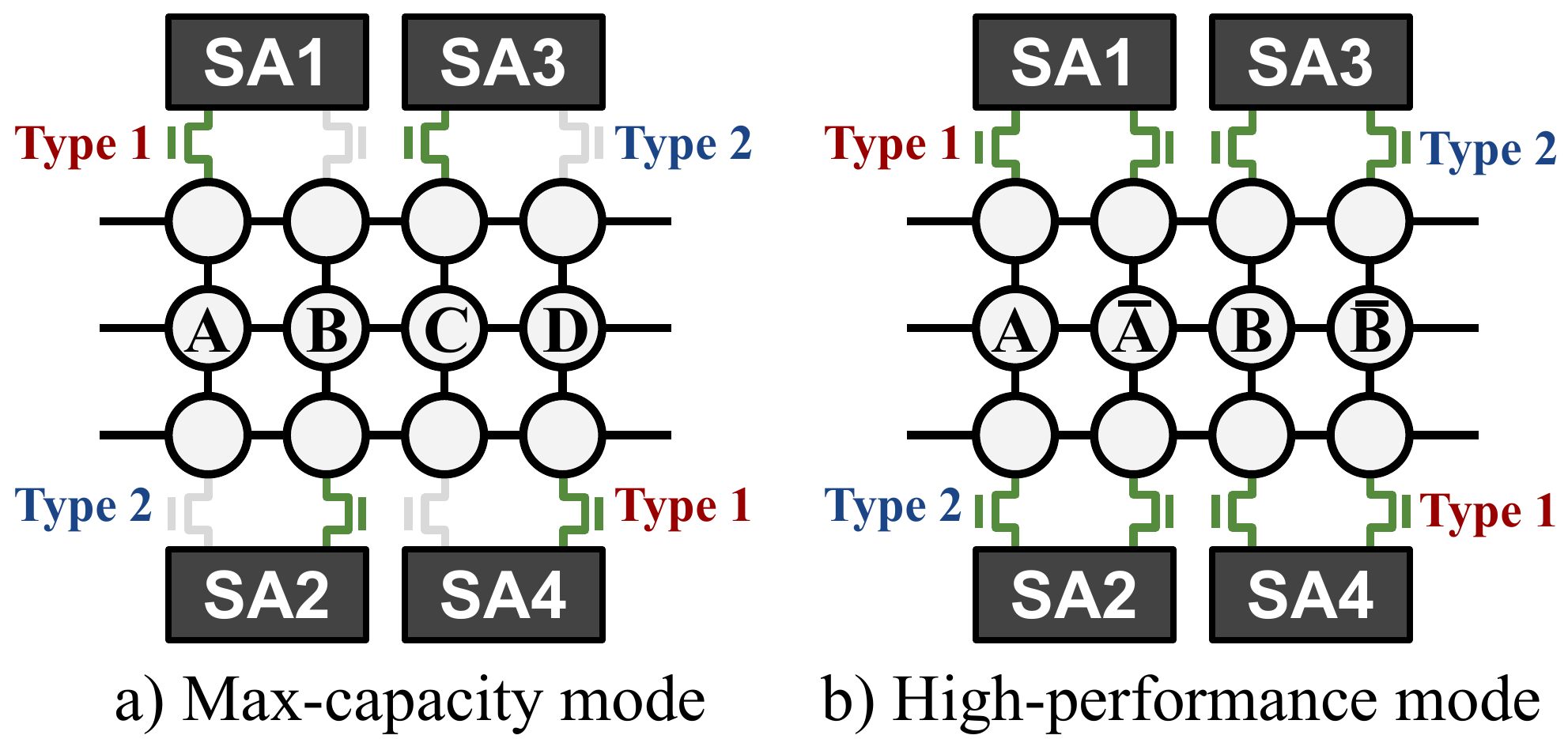}
\caption{Max-capacity mode vs. high-performance mode: connections of bitline mode select transistors}
\label{fig:TCTS_Control}
\end{figure}

 \mechanism{} enables a DRAM row to operate in either max-capacity mode or high-performance mode by reconfiguring the connections between DRAM cells and SAs using the two types of bitline mode select transistors (\textbf{Type 1} and \textbf{Type 2}). \mechanism{} can reconfigure such connections when a DRAM row is accessed (i.e., activated). Therefore, the operating mode of a row is independent from that of any other row. Figure~\ref{fig:TCTS_Control} shows how  \textbf{Type 1} and \textbf{Type 2} bitline mode select transistors are configured to operate in max-capacity and high-performance modes, for an example subarray with three rows and four cells per row. A green (faded gray) transistor indicates that the transistor is enabled (disabled).




\par In max-capacity mode (Figure~\ref{fig:TCTS_Control}a), \mechanism{} adopts the open-bitline DRAM architecture's (Figure~\ref{fig:TCTS_Arch}a) cell-to-SA connections to achieve the same storage capacity as in the baseline open-bitline DRAM architecture. 
In this mode, \mechanism{} connects each of the adjacent DRAM cells in a row (e.g., cells \textbf{A} and \textbf{B} in Figure~\ref{fig:TCTS_Control}a) to a \emph{different} SA on either side of the subarray (cell \textbf{A} to \textbf{SA1} and cell \textbf{B} to \textbf{SA2}) to make each DRAM cell operate independently of the other. When accessing a row in max-capacity mode,
\mechanism{} enables \textbf{Type 1} transistors and disables \textbf{Type 2} ones in the subarray.
 

\par In high-performance mode, \mechanism{} \emph{simultaneously} couples each of the 1) adjacent cells in a row (e.g., cell \textbf{A} and $\overline{\textbf{A}}$ in Figure~\ref{fig:TCTS_Control}b) to operate as a single \emph{logical cell} with high effective capacitance, and 2) corresponding two SAs (e.g., \textbf{SA1} and \textbf{SA2}), which drive the coupled cells individually in max-capacity mode, to operate as a \emph{single} but \emph{stronger} logical SA driving the logical cell. In this mode, \mechanism{} reconfigures the connections between DRAM cells and SAs to achieve coupled operation of adjacent cells and their two corresponding SAs in two ways (Figure~\ref{fig:TCTS_Control}b). First, to couple two adjacent DRAM cells as a logical cell, \mechanism{} connects \emph{both} cells to the \emph{same} SA. For example, cell \textbf{A} is connected to the $bitline$ port of \textbf{SA1}, and cell $\overline{\textbf{A}}$ is connected to the $\overline{bitline}$ port of \textbf{SA1}. Second, to couple two SAs as a logical SA, \mechanism{} connects the logical cell to \emph{both} of the SAs. For example, the logical cell formed by the coupled cells \textbf{A} and $\overline{\textbf{A}}$ is connected to both \textbf{SA1} on the top and \textbf{SA2} at the bottom of the subarray. When accessing a row in high-performance mode, \mechanism{} enables both \textbf{Type 1} and \textbf{Type 2} transistors in the subarray.

 \par
 Although this coupled operation of adjacent DRAM cells and their two sense amplifiers sacrifices half of a row's total storage capacity, it improves DRAM cell access performance in three ways. First, it significantly reduces cell access latency by increasing the total effective cell capacitance and strengthening the overall sense-amplification capabilities (Section~\ref{Operation}). Second, it enables early-termination of the charge restoration process to further reduce the row activation latency by exploiting the difference in charge restoration speed between the coupled discharged and charged cells (Section~\ref{earlyprecharge}). Third, it reduces the cost of refresh operations by enabling a lower refresh latency and a longer retention time (Section~\ref{reducerefresh}). 


 \subsection{Control of Bitline Mode Select Transistors}\label{Control}

\par


\par To reconfigure the connections between DRAM cells and SAs \emph{within} a subarray as described in Section~\ref{twomodes} for max-capacity and high-performance operation, we introduce two control signals, \texttt{ISO1} and \texttt{ISO2}\mpg{,} to control the \textbf{Type 1} and \textbf{Type 2} bitline mode select transistors in the subarray, respectively. For example, asserting \texttt{ISO1} and deasserting  \texttt{ISO2} configures a row to operate in max-capacity mode, and asserting both \texttt{ISO1} and \texttt{ISO2} configures a row to operate in high-performance mode. 

\par 
At the same time, the control circuitry of  \mechanism{} must properly configure the bitline-SA connections in the two \emph{neighboring} subarrays to: 1) ensure correct operation of the SAs in max-capacity mode by \emph{connecting} certain bitlines in the \emph{adjacent} subarrays to the SAs as in the conventional open-bitline architecture (Figure~\ref{fig:TCTS_Arch}a), and 2) maximize the latency reduction in high-performance mode by \emph{disconnecting} \emph{all} bitlines in the \emph{neighboring} subarrays from the SAs to avoid increasing the effective bitline length (and hence the capacitance seen by the SAs).

\par
To satisfy the aforementioned requirements \emph{without} introducing extra control signals, \mechanism{} alternates between applying two sets of control signal assignments  \{$\texttt{ISO1}\xrightarrow{}\textbf{Type 1}$, $\texttt{ISO2}\xrightarrow{}\textbf{Type 2}$\} and  \{ $\overline{\texttt{ISO2}}\xrightarrow{}\textbf{Type 1}$, $\overline{\texttt{ISO1}}\xrightarrow{}\textbf{Type 2}$ \} to odd and even numbered subarrays, respectively.
\par
\begin{figure}[h]
\centering
\includegraphics[width=0.5\textwidth]{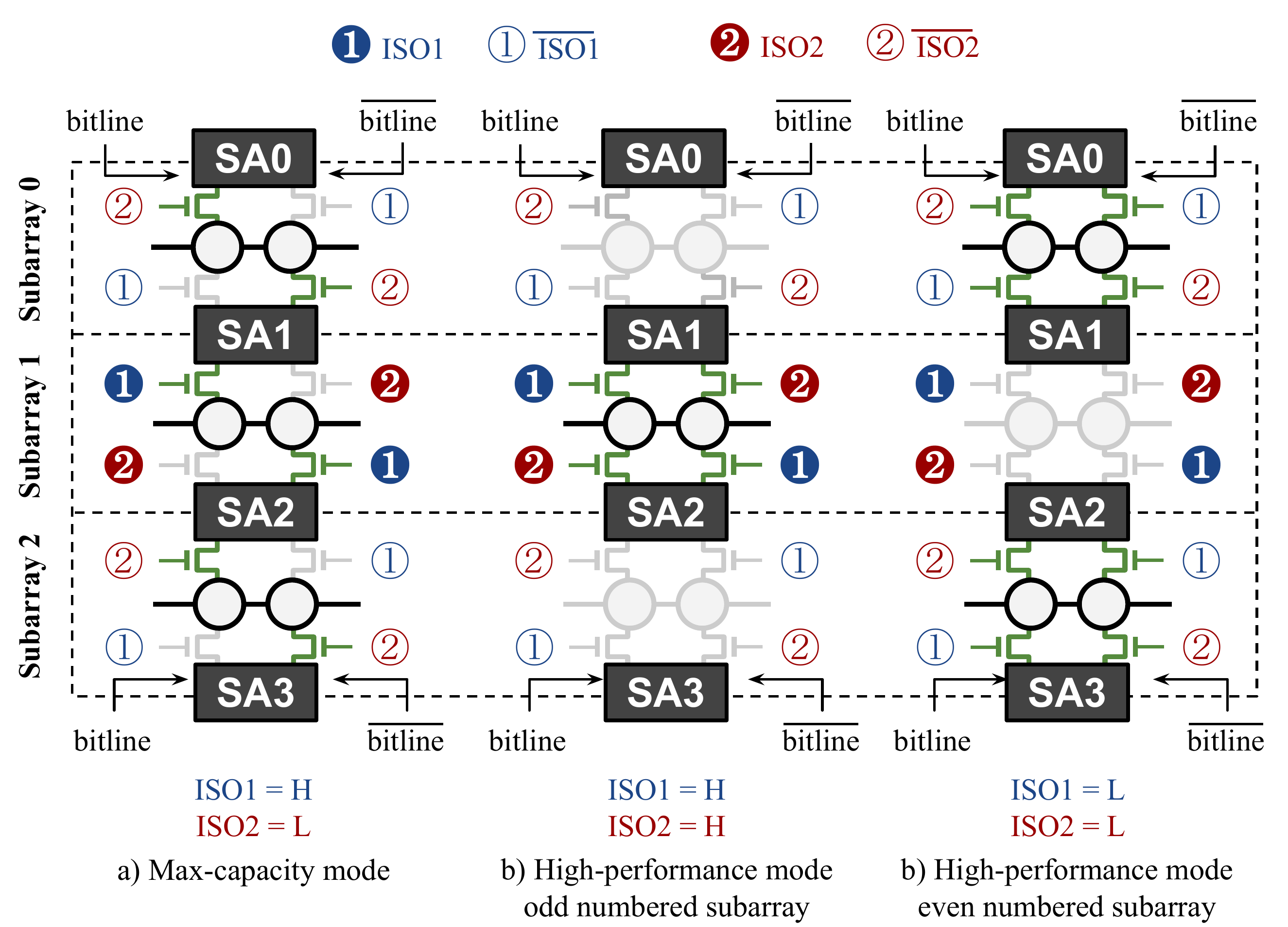}
\caption{Bitline mode select transistor control in adjacent subarrays.}
\label{fig:TCTS_Control_neighbor}
\end{figure}

Figure~\ref{fig:TCTS_Control_neighbor} illustrates \mechanism{}'s operation in max-capacity and high-performance modes with the previously-mentioned alternating control signal assignments across adjacent subarrays. We simplify our explanations by depicting only one row consisting of two adjacent DRAM cells in each subarray. Enabled transistors are marked in green.

\par 
To configure a row in either an odd-numbered or even-numbered subarray to operate in max-capacity mode, \mechanism{} asserts $\texttt{ISO1}$ and deasserts $\texttt{ISO2}$ (Figure~\ref{fig:TCTS_Control_neighbor}a). Doing so configures the connections between DRAM cells and SAs to be equivalent to the conventional Open-Bitline DRAM architecture, where cells and SAs operate individually.

\par 
To configure a row in an odd-numbered subarray (e.g., \textbf{Subarray 1} in Figure~\ref{fig:TCTS_Control_neighbor}b) to operate in high-performance mode, \mechanism{} \emph{asserts} both $\texttt{ISO1}$ and $\texttt{ISO2}$. For an even-numbered subarray (e.g.,  \textbf{Subarray 0} or \textbf{Subarray 2} in Figure~\ref{fig:TCTS_Control_neighbor}c), \mechanism{} \emph{deasserts} both $\texttt{ISO1}$ and $\texttt{ISO2}$. In both cases, all bitline mode select transistors in the subarray are enabled to couple every two adjacent DRAM cells and their two SAs in the row to operate as a single logical cell driven by a single logical SA. By assigning the \emph{complements} of the control signals to neighboring subarrays, all bitlines in neighboring subarrays are \emph{disconnected} to prevent them from degrading the latency reduction benefits in high-performance mode.

\subsection{Reducing Access Latency in High-Perf. Mode} 
\label{Operation}


\par Operating in high-performance mode changes 
how the cells and the sense amplifiers are connected to achieve two kinds of coupled operation \emph{at the same time}. First, high-performance mode couples every two adjacent physical DRAM cells in a row to simultaneously operate as a single logical cell. It does so by connecting the two cells to both the \emph{\bl{}} and \emph{\blbar{}} ports of a sense amplifier. Second, high-performance mode couples two sense amplifiers to operate as a single logical sense amplifier driving the single logical cell. \mechanism{} reduces access latency because the single logical cell is simultaneously connected to two sense amplifiers, which originally served the two cells individually (i.e., in the baseline architecture or in max-capacity mode), on both sides of a subarray. The two coupled sense amplifiers drive the single logical cell from \emph{both ends} of the same pair of bitlines.

\par
\begin{figure}[!h]
\centering
\includegraphics[width=0.47\textwidth]{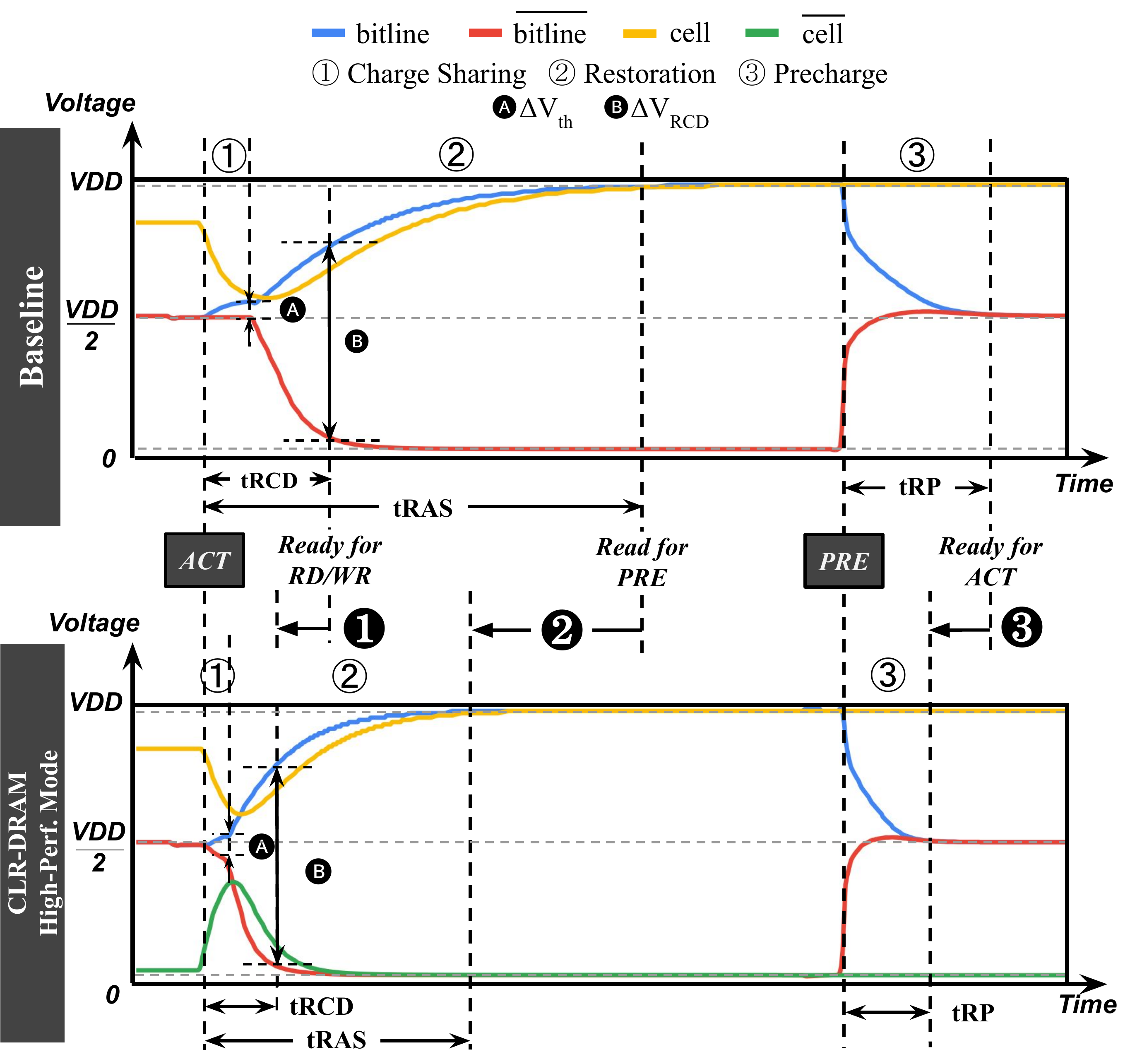}
\caption{SPICE simulation of row activation and precharge of  baseline (top) and \mechanism{} high-performance mode (bottom).}
\label{fig:TCTS_OP}
\end{figure}

 To illustrate the detailed behavior of high-performance mode, Figure~\ref{fig:TCTS_OP} shows our  SPICE simulation of the change in voltage level on the pair of bitlines (\emph{\bl{}} and \emph{\blbar{}}) and the two coupled cells (\emph{cell} and $\overline{cell}$) during row activation and precharge in \mechanism{}'s high-performance mode (bottom), and the baseline open-bitline architecture (top). We make three observations. 

\par First, two coupled cells always store opposite charge levels (i.e., charged and discharged) to represent the same logical bit (\emph{cell} and $\overline{cell}$  in Figure~\ref{fig:TCTS_OP} (bottom)), because the sense amplifiers always drive \emph{\bl{}} and \emph{\blbar{}} to complementary voltage levels.

\par Second, two coupled cells with opposite charge levels drive their respective bitlines in opposite directions during the initial charge sharing phase of a row activation \hcircled{1}. This creates a larger $\Delta V$ across \emph{\bl{}} and \emph{\blbar{}} \circled{A} relative to the baseline architecture, where only one cell drives either \emph{\bl{}} or \emph{\blbar{}}, but \emph{not both}. Because a sense amplifier enters the charge-restoration phase \hcircled{2} when $\Delta V$ reaches the threshold level $\Delta V_{th}$, a larger $\Delta V$ causes the threshold level to be reached more quickly. By shortening the charge sharing phase, we reduce
DRAM timing parameters that depend on the charge sharing latency (i.e., \textbf{tRCD} \circled{1} and \textbf{tRAS} \circled{2}).

\par Third, the reduction in \textbf{tRCD} and \textbf{tRAS} is larger than the reduction in the charge sharing time. This is because coupling two sense amplifiers to drive the two coupled cells simultaneously from both ends of the bitline pair \emph{fundamentally} accelerates the entire row activation process. As explained in Section~\ref{sec:organization}, coupling two sense amplifiers also couples the two precharge units associated with each of them, thereby also reducing the precharge latency (\textbf{tRP} \circled{3}).

\subsection{Early-Termination of Charge Restoration}
\label{earlyprecharge}

\mechanism{} reduces the latency of DRAM operations involving charge restoration  (e.g., \textbf{tRAS} of row activations)  for high-performance mode rows even further (on top of reductions shown in Section~\ref{Operation}) by exploiting three observations on the charge restoration process of coupled cells. 
\par First, the two coupled cells storing opposite levels of charge have higher effective capacitance compared to a single cell. This allows the cell to tolerate more charge leakage while still maintaining the target retention time (e.g., 64ms). 

\par
\begin{figure}[!h]
\centering
\includegraphics[width=0.47\textwidth]{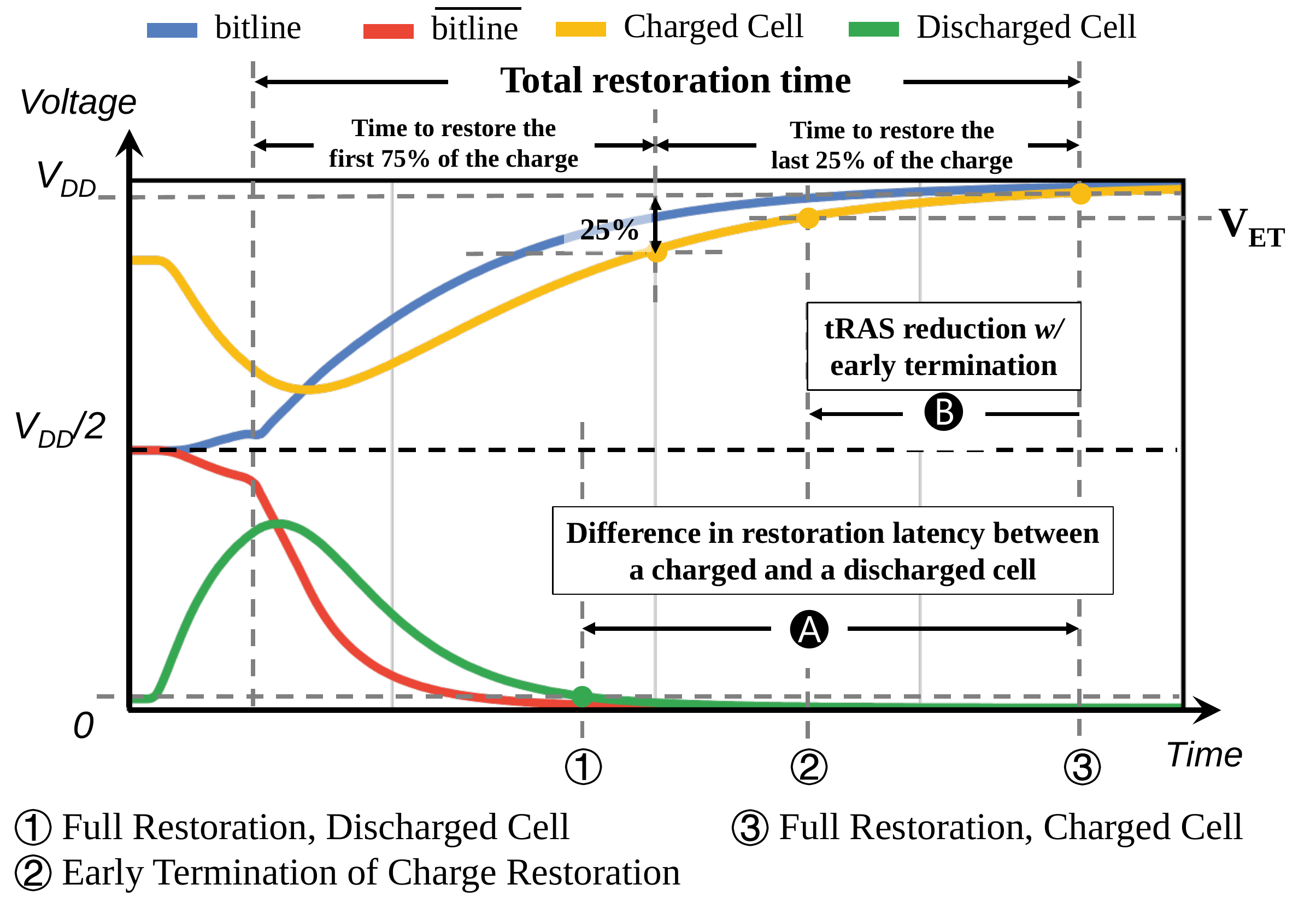}
\caption{Reduction in \textbf{tRAS} enabled by terminating charge restoration early.}
\label{fig:TCTS_ET_waveform}
\end{figure}

 Second, the charge restoration phase has a long tail latency. As shown in our example SPICE simulation of a row activation (Figure~\ref{fig:TCTS_ET_waveform}), about half of the charge restoration phase is spent on restoring the last 25\% of charge to the charged cell. In other words, terminating the charge restoration process before the cell is fully restored to $V_{DD}$ does \emph{not} significantly degrade the charge level in the cell. 

\par Third, the charge restoration speed of a discharged cell is \emph{fundamentally} faster than that of a charged cell (Figure~\ref{fig:TCTS_ET_waveform} \circled{A}). The reason for this difference is similar to that of the difference in charge-sharing speed between charged and discharged cells~\cite{keeth2007dram}, resulting in asymmetric charge restoration currents for charged and discharged cell capacitors. This means that we can terminate charge restoration \emph{before} the charged cell is fully restored but \emph{after} the discharged cell is, and still reduce  the charge sharing time.

Leveraging the three aforementioned observations, we terminate the charge restoration phase of the row activation \emph{early} when the \emph{charged} cell in the two coupled cells reaches $V_{ET}$ (early termination voltage level), a level smaller than the conventional full-restoration level ($V_{DD}$), to further reduce \textbf{tRAS} (Figure~\ref{fig:TCTS_ET_waveform} \circled{B}). We set $V_{ET}$ to be small enough to provide a non-trivial reduction of \textbf{tRAS} while ensuring that the following two requirements are met. First, the coupled cell must still meet the target retention time (e.g., 64\,ms) when charge restoration is terminated early. Second, $V_{ET}$ should not be so small that it significantly degrades the amount of charge available for the next row activation, which could in turn require increasing \textbf{tRCD}.
Our analysis (Section~\ref{SPICE}) shows that early termination applies to both row activation and write restoration, whose charge restoration phases are analogous to each other. By terminating charge restoration early for both operations, we reduce \textbf{tRAS}  and \textbf{tWR} significantly (by more than 30\%)  on top of the \textbf{tRAS} reductions provided by techniques described in Section~\ref{Operation}, while only marginally increasing \textbf{tRCD} (by less than 2\%) due to reduction in the amount of charge available for the next activation.

\subsection{Mitigating Refresh Costs}
\label{reducerefresh}
\par \mechanism{} mitigates the performance and energy overheads of refresh in two ways: 1) reducing refresh latency and 2) reducing refresh rate.

\par \textbf{Reducing Refresh Latency.} Since a refresh operation is essentially an activation followed by a precharge, whose latencies are both reduced in \mechanism{}'s high-performance mode, \mechanism{} also reduces the latency of a refresh operation (i.e., refresh cycle time, \textbf{tRFC}) for a row operating in high-performance mode.

\par \textbf{Reducing Refresh Rate.} The logical cell formed by coupling two adjacent DRAM cells has a larger capacitance compared to a single DRAM cell. In other words, a logical cell can tolerate \emph{more} leakage than a single DRAM cell without affecting the integrity of the data stored in it, which increases its retention time and thus allows reducing the refresh rate (i.e., increasing the refresh window \textbf{tREFW}).
\par Extending \textbf{tREFW} limits how much \mechanism{} can reduce the row activation latency (\textbf{tRCD} and \textbf{tRAS}) in high-performance mode. A larger \textbf{tREFW} essentially reduces the charge level of the logical cell prior to  activation, requiring a longer charge sharing phase for the SA to correctly sense the stored data. Section~\ref{SPICE-sensitivity} provides a sensitivity analysis quantifying the trade-off between the refresh window (\textbf{tREFW}) and the row activation latency (\textbf{tRCD} and \textbf{tRAS}) in high-performance mode.
\section{CLR-DRAM Column Access}

We design \mechanism{}'s column I/O circuitry to maintain 
full column access  bandwidth between the SAs and the global I/O circuitry
in both max-capacity and high-performance modes. To do so, we add a single isolation transistor, called \emph{column I/O mode select transistor}, for each pair of SAs.
It enables \mechanism{} to connect \emph{only one} of two coupled SAs to the global I/O circuitry in high-performance mode to avoid transferring duplicated data values from the coupled SAs.




\par
\begin{figure}[h!]
\centering
\includegraphics[width=0.47\textwidth]{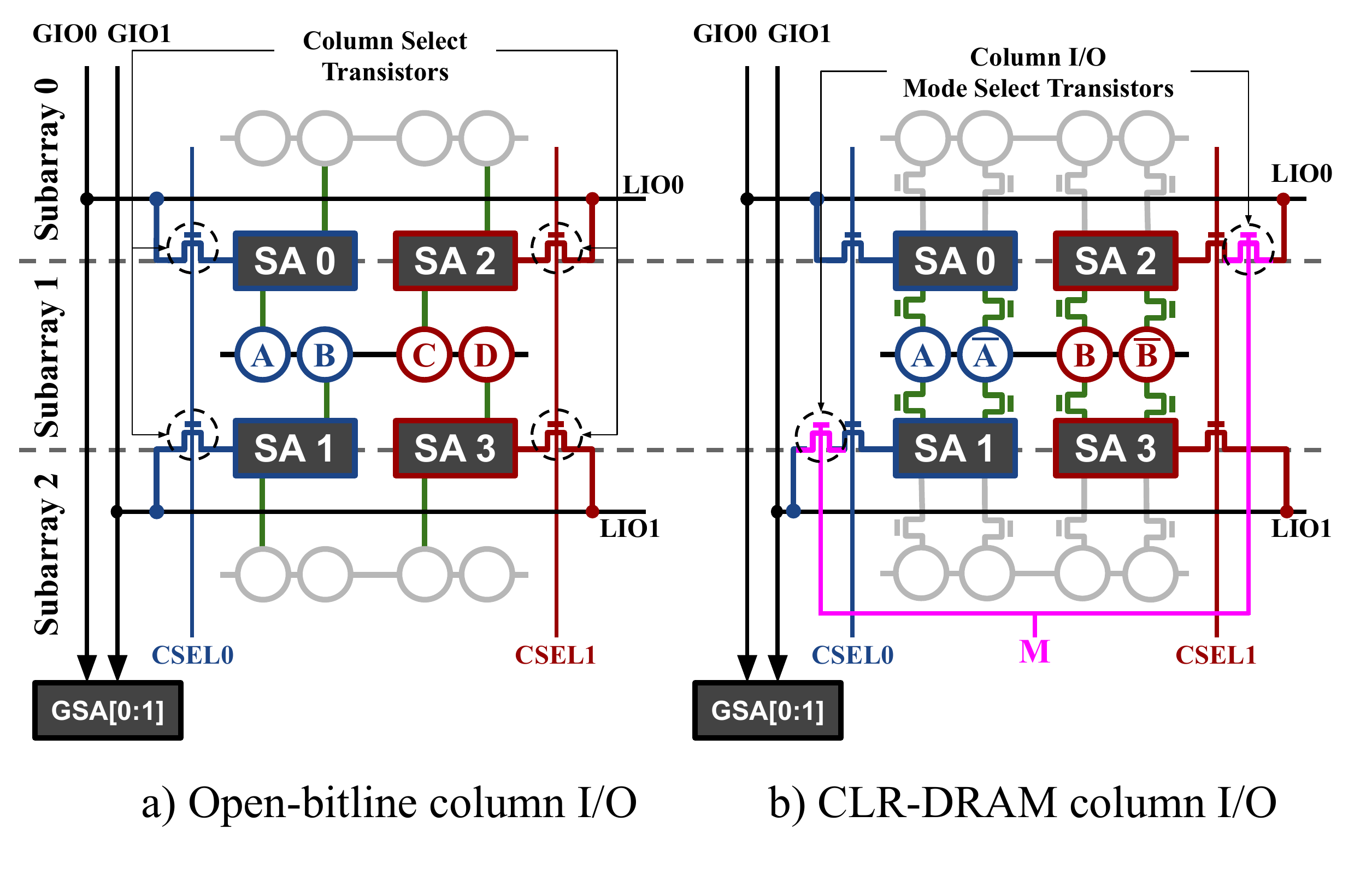}
\caption{Subarray column I/O circuitry.}
\label{fig:Base_IO}
\end{figure}

 \textbf{Conventional Column Access.} Figure~\ref{fig:Base_IO}a illustrates a simplified conventional column I/O architecture, 
in which a row contains only two columns, and each column contains two cells (e.g., cells \textbf{A} and \textbf{B} form the blue column, cells \textbf{C} and \textbf{D} form the red column). 
The I/O circuitry is responsible for communicating a \emph{column} of data (e.g., bits in cell \textbf{A} and \textbf{B}) from a subarray's local SAs (e.g., \textbf{SA0} and \textbf{SA1}) to the bank-level global SAs (e.g., \textbf{GSA[0:1]}) via \emph{local} I/O (e.g., \jsc{9}{\textbf{LIO0}} and \jsc{9}{\textbf{LIO1}}) and 
\emph{global} I/O (e.g., \jsc{9}{\textbf{GIO0}} and \jsc{9}{\textbf{GIO1}}) lines. Column-select signals (e.g., \jsc{9}{\texttt{CSEL0}} and \jsc{9}{\texttt{CSEL1}}) control column select transistors to ensure that at most one column in a subarray is connected to LIO lines at a given time. Similarly, GIO lines are time-multiplexed across subarrays in a bank. 

\par \textbf{CLR-DRAM Column Access.} Since \mechanism{}'s high performance mode couples two sense amplifiers of two adjacent cells (e.g., \jsc{9}{\textbf{SA0}} and \jsc{9}{\textbf{SA1}} in Figure~\ref{fig:Base_IO}b) as one logical sense amplifier, employing the conventional column I/O architecture would waste half of the available LIO and GIO lines to transfer redundant bits (e.g., both \jsc{9}{\textbf{SA0}} and \jsc{9}{\textbf{SA1}} contain data ``\jsc{9}{\textbf{A}}''). To fully utilize the available column I/O bandwidth (as in the conventional subarray architecture and max-capacity mode), \mechanism{} adds an isolation transistor, called the \emph{column I/O mode select transistor}, for each pair of physical sense amplifiers (as shown in Figure~\ref{fig:Base_IO}b).
We use a \emph{column I/O mode select signal \jsc{9}{\texttt{M}}} that controls \emph{all} column I/O mode select transistors in a subarray. In max-capacity mode, \jsc{9}{\texttt{M}} is asserted so that the subarray I/O operates in a way equivalent to the conventional subarray I/O.
In high-performance mode, \jsc{9}{\texttt{M}} is deasserted to disconnect the \emph{different} redundant halves of the columns (e.g., \textbf{SA1} in the blue column and \textbf{SA2} in the red column) from the I/O circuitry, while asserting two column select signals (e.g., \jsc{9}{\texttt{CSEL0}} and \jsc{9}{\texttt{CSEL1}}) to utilize all of the available data bandwidth. For example, \mechanism{} can simultaneously access two \emph{different} bits in \textbf{SA0} and \textbf{SA3} in Figure~\ref{fig:Base_IO}b in high-performance mode.

\section{System Integration}
\label{sec:system_integration}

\mechanism{} enables a dynamic system-level trade-off between DRAM capacity and DRAM latency adjustable at the granularity of a DRAM row. The system is free to expose this trade-off to any level of the system stack. However, there are two challenges that may complicate communication between \mechanism{} and other parts of the system: 1) controlling the data mapping between physical addresses and DRAM rows due to memory controller address mapping policies and 2) DRAM refresh control logic. This section discusses each of these challenges and explores potential solutions.

\subsection{Memory Controller Address Mapping}
\label{addrinterleaving}
\par
The memory controller typically applies an address interleaving policy that distributes memory accesses throughout different levels of the DRAM hierarchy (i.e., channels, ranks, banks) to improve memory access performance by exploiting parallelism in DRAM~\cite{pessl2016drama, seaborn2015physical, bkdg_amd2013, intel2007intel, intel202010th, kim2010atlas, suzuki2013coordinated, kim2015ramulator, ramulatorgithub}.
The address interleaving policy can potentially split up a single physical page so that different parts of the page are mapped to different DRAM rows.
Therefore, \emph{multiple} DRAM rows may \emph{have to} be configured to operate in \emph{high-performance mode} when software requests the allocation of a single low-latency memory page through the OS.
As a consequence, because each DRAM row may contain parts of multiple different physical pages, all of the physical memory pages that the newly-configured high-performance rows collectively contain also become low-latency pages. Thus, the address interleaving policy can increase the granularity at which the system interfaces with \mechanism{}.

\par
\begin{figure}[h!]
\centering
\includegraphics[width=0.48\textwidth]{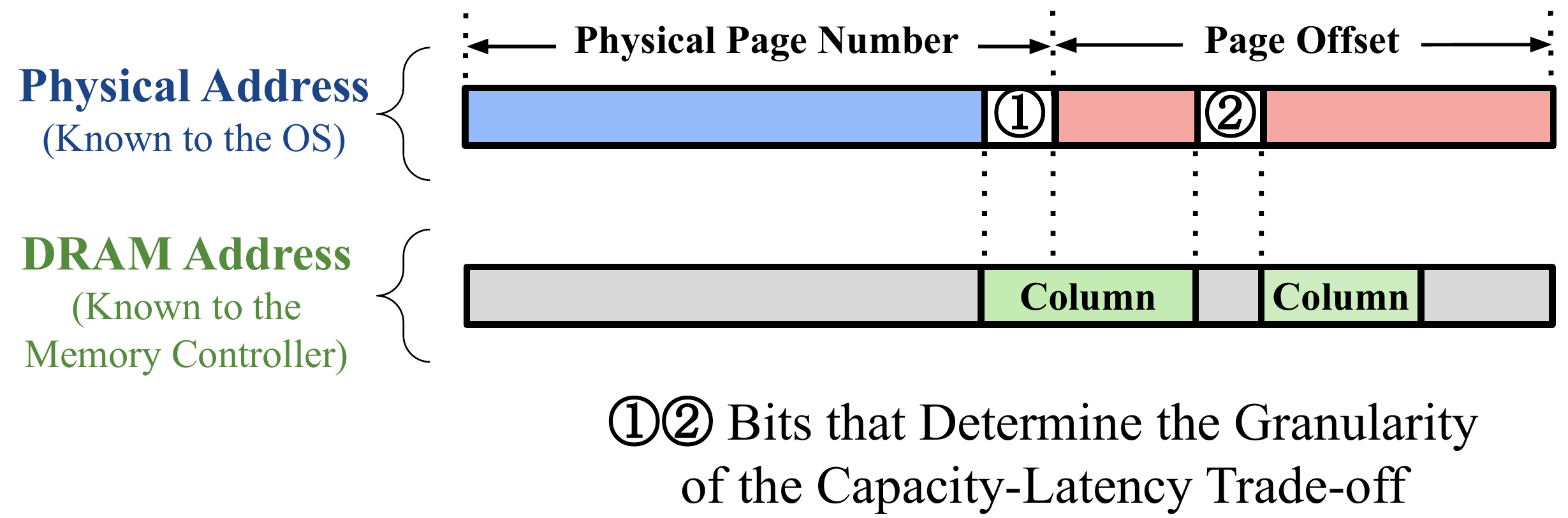}
\caption{Physical address to DRAM address mapping.}
\label{fig:addrmap}
\end{figure}

 Figure~\ref{fig:addrmap} provides an example address interleaving scheme and shows how the memory controller might map physical addresses to different levels of the DRAM hierarchy. 
The memory controller generates the DRAM column address using 
the bits marked \hcircled{1}, taken from the \emph{page number} field of the physical address, but \emph{not} the bits marked \hcircled{2}, taken from the \emph{page offset} field of the physical address.
Thus, for a given physical address, bits \hcircled{1} indicate how many other pages are also mapped to the same DRAM row and bits \hcircled{2} indicate how many different DRAM rows the page is striped across.



\par For example, if \hcircled{1} has $X$ bits and \hcircled{2} has $Y$ bits, when the system specifies that a page should have low latency, a total of $\frac{1}{2}\cdot 2^{X}$ pages\footnote{High-performance rows provide \emph{half} of the total row capacity, so we multiply the result by $\frac{1}{2}$.} will become low-latency pages (since the entirety of the row will be in high-performance mode), causing $2^{Y}$ rows in \mechanism{} to switch to high-performance mode.


\par To best take advantage of such a coarse reconfiguration granularity, we believe that the system should ``gang together'' multiple pages that would benefit from being mapped to a high performance row and map them together. We believe such information about multiple pages can be communicated from the software to hardware via changes to the system software and can be done more easily by adopting new frameworks such as VBI~\cite{hajinazar2020virtual}, Labeled RISC-V~\cite{yu2017labeled}, PARD~\cite{ma2015supporting}, and XMem~\cite{vijaykumar2018case}. We leave the exploration of the (system) software stack for CLR-DRAM to future work.

\subsection{Supporting Heterogeneity in Refresh}
\label{Refcontrol}

\par High-performance mode rows in \mechanism{} introduce heterogeneity in DRAM refresh operations in two ways. First, compared to rows operating in max-capacity mode, the latency of a refresh operation (\textbf{tRFC}) is reduced due to the faster row activation and precharge. Second, the time interval between two consecutive refresh operations to a high-performance row (\textbf{tREFW}) can be extended compared to a max-capacity row. Such heterogeneity requires the refresh control logic to be able to distinguish between rows operating in max-capacity mode and high-performance mode to 1) apply different \textbf{tRFC} timing constraints, and 2) issue refresh commands at different rates. 

There are many solutions proposed by prior works to efficiently handle heterogeneity in DRAM refresh operations~\cite{liu2011flikker, liu2012raidr, cui2014dtail, chang2014improving, qureshi-dsn2015, bhati2015flexible, das2018vrldram, patel2017reach, khan2016case, khan-sigmetrics2014, khan2016parbor,khan2017detecting}. We believe these techniques and designs could be adapted to \mechanism{}.
\section{Capacity and Hardware Overhead Analysis}
\label{overhead}
We analyze the capacity and hardware (Sections~\ref{subsec:capacity_overhead} and \ref{subsec:hardware_overhead}) overheads that \mechanism{} introduces, including changes to the subarray, the subarray column I/O circuitry, and the memory controller.

\subsection{DRAM Capacity Overhead}
\label{subsec:capacity_overhead}
\par Any DRAM row configured to operate in high-performance mode provides half of the storage capacity of a conventional DRAM row. Therefore, configuring $X$\% of all rows in a DRAM bank to operate in high-performance mode results in an $\frac{X}{2}$\% reduction in the total DRAM capacity. We argue that this reduction in memory capacity is not a significant downside of \mechanism{} due to two reasons.

\par First, \mechanism{} allows the user or the system to \emph{dynamically} reconfigure DRAM and find a favorable operating point in the DRAM capacity-performance trade-off space. In edge cases where the system requires more memory capacity (e.g., to avoid frequent page swaps), it can configure more DRAM rows in \mechanism{} to operate in max-capacity mode.

\par Second, previous works show that many modern workloads under-utilize the main memory capacity (e.g., in HPC~\cite{Panwar2019MemUtil}, cloud~\cite{Reiss2012GoogleCloud, Chen2018Alibaba}, and enterprise~\cite{di2012characterization} domains).
In such workloads, the user or system can trade the under-utilized memory capacity to improve system performance by switching more DRAM rows to high performance mode.

We leave it to the user or the system software to determine a suitable balance between the two operating modes. A user cognizant of a workload's memory usage characteristics can use \mechanism{} to configure DRAM rows to operate in high-performance or max-capacity mode as needed by the workload. We leave a rigorous exploration of how different workloads and systems can exploit the new dynamic capacity-latency reconfigurability and the associated trade-offs enabled by \mechanism{} to future work.


\subsection{Hardware Overhead}
\label{subsec:hardware_overhead}

\mechanism{} requires modest hardware changes to the DRAM chip and memory controller circuitry. This section analyzes the impact of these changes on overall chip area.


\noindent\textbf{DRAM Chip Overhead.} 
\mechanism{} adds two sets of isolation transistors to a density-optimized DRAM chip: 1) \emph{bitline mode select transistors} and 2) \emph{column I/O mode select transistors}.
First, two bitline mode select transistors added to each bitline in a subarray (i.e., one at either end of each bitline) enable switching of a row between the two operating modes. 
 This increases the height of a DRAM subarray, resulting in an area overhead of 1.6\% of the DRAM chip.\footnote{We size the isolation transistors according to prior work~\cite{Seongil2014RBD,ptmweb}.} Second, \mechanism{} adds column I/O mode select transistors next to half of the existing column select transistors to provide the same column I/O data transfer bandwidth in both high-performance and max-capacity modes.
While it is possible that the column I/O mode select transistors may fit into existing slack space without increasing the chip area, we conservatively assume that no slack space is available for \mechanism{} to exploit.\footnote{We assume the worst case due to the lack of open literature on modern DRAM I/O circuitry designs that we could use to accurately evaluate whether the mode select transistors can fit into the existing slack area.} 
 Assuming that a column I/O mode select transistor requires the same area as a bitline mode select transistor, it incurs an additional 1.6\% DRAM chip area overhead.
Overall, even with this conservative estimate,  
\mechanism{} increases the DRAM chip area by only 3.2\%, compared to a commodity density-optimized DRAM baseline.


\noindent\textbf{Memory Controller Overhead.} In \mechanism{}, the memory controller uses two separate sets of timing parameters to access and refresh DRAM rows in max-capacity mode and high-performance mode. 
To fully utilize the row-level reconfigurability, the memory controller needs to track the operating mode of each DRAM row to apply proper timing parameters. Without optimization, this requires storing one bit per row. However, the required storage reduces by a factor of $2^{Y}$ if the memory controller address mapping requires the reconfiguration granularity of \mechanism{} to be larger than a single row (discussed in Section~\ref{addrinterleaving}).
The required storage may be optimized even further using a more efficient representation. Many space-efficient data structures are commonly used in hardware systems (e.g., sparse bitmaps~\cite{kanellopoulos2019smash}), and systems integrating \mechanism{} can potentially make use of similar approaches.


\subsection{Handling Row and Column Redundancy}
\par A modern DRAM chip internally remaps known faulty DRAM rows/columns to redundant rows/columns that are provided as spares to handle post-manufacturing faults~\cite{horiguchi2011nanoscale}.
Because \mechanism{} performs reconfiguration at the granularity of a single row, 
it is fully compatible with existing row redundancy resources. However, our design may use \emph{more} redundant columns compared to conventional DRAM because, if a faulty column is remapped to a spare column, its adjacent column must also be remapped to the corresponding adjacent spare column in order to ensure that all cells throughout the row can be coupled pairwise for high-performance mode operation. For commodity devices in the field, redundant rows and columns are typically underutilized~\cite{kim2016ecc, ooishi1998synchronous} (e.g., <25\% utilization~\cite{kim2016ecc}). Therefore, \mechanism{} likely does not require increasing the amount of available redundant row or column resources to implement the high-performance mode.

\section{Circuit Simulations}

We use SPICE circuit simulations to evaluate the latency and refresh reductions \mechanism{} provides.

\subsection{Methodology}
We use a methodology similar to that of prior work~\cite{Hassan2019CROW}. We model a DRAM subarray in SPICE based on Rambus DRAM technology parameters~\cite{rambus} and scale the technology parameters to 22\,nm according to the ITRS roadmap~\cite{itrs, vogelsang2010understanding}.
We have open sourced our SPICE circuit model~\cite{clrdram}.
\begin{itemize}[leftmargin=*]
    \item We model the sense amplifier using the 22nm PTM high-performance transistor model~\cite{ptmweb}.

    \item We model the worst-case operating conditions with a temperature of 85\textdegree{}C.
    
    \item We assume the junction leakage towards the body of the access transistors is the major charge leakage path of the cell \cite{Lee2011SimultaneousRB, takemura20060,Itoh95TrendDRAM,Sim04CD,Tanaka99NWL,Tajalli09SubtReduction}.
    
    \item We tune other parameters (e.g., bitline capacitance and resistance) to the best of our ability so that the timing parameters derived from our model closely approximate real DDR4 datasheet values~\cite{jedec-ddr4}.
\end{itemize}

\par We model manufacturing process variation by running $10^4$ iterations of Monte Carlo simulations with 5\% variation in every circuit component. We derive the timing parameters of our design based on the slowest of the $10^4$ iterations and make sure that every single iteration reads the correct value from the the cell.

\subsection{Latency Reduction}
\label{SPICE}
\hh{Table~\ref{tab:timing_overview} shows the key DRAM timing parameters of conventional DRAM (i.e., our baseline) and the two operating modes of \mechanism{}. }
Overall, we observe 35--65\% reduction in the four key timing parameters in high-performance mode.

\begin{table}[h]
\small
\setlength\tabcolsep{2pt}
\centering
\vspace*{.07in}
\caption{Reduction in major DRAM timing parameters.}
\begin{tabular}{c||c||c|c|c||c}
\hline
\multirow{2}{*}{\begin{tabular}[c]{@{}c@{}}Timing\\ Parameter\end{tabular}} & \multirow{2}{*}{Baseline} & \multirow{2}{*}{Max-Cap.} & \multicolumn{2}{c||}{High-Performance} & Reduction \\ \cline{4-5}
 &  &  & \multicolumn{1}{c|}{w/o E.T.} & \multicolumn{1}{c||}{w/ E.T.} & (w/ E.T.) \\ \hline\hline
\textbf{tRCD (ns)} & 13.8 & 13.2 & 5.4 & 5.5 & \textbf{60.1\%} \\ \hline
\textbf{tRAS (ns)} & 39.4 & 40.3 & 20.3 & 14.1 & \textbf{64.2\%} \\ \hline
\textbf{tRP (ns)} & 15.5 & \multicolumn{3}{c||}{8.3} &\textbf{46.4\%}
\\ \hline
\textbf{tWR (ns)} & 12.5 & 13.3 & 12.5 & 8.1 & \textbf{35.2\%} \\ \hline
\end{tabular}

\label{tab:timing_overview}
\end{table}

\par \noindent\textbf{Impact of Bitline Mode Select Transistors in Max-Capacity Mode.} The timing parameters for rows operating in max-capacity mode are different from those of the baseline because of the insertion of the bitline mode select transistors. The bitline mode select transistors impact timing in three ways. First, they \emph{slightly} reduce \textit{\textbf{tRCD}} by 4.4\% because they decouple the SAs from the long bitlines that have large capacitance. Second, at the same time, they also limit the amount of current that can pass through the bitline, making \textit{\textbf{tRAS}} and \textit{\textbf{tWR}} \emph{slightly} higher than in the baseline (by 2.2\% and 6.4\%, respectively). 
Third, they enable \mechanism{} to reduce the precharge latency \emph{\textbf{tRP}} by 46.4\% \emph{regardless of} the target row's operating mode, by coupling \emph{two} precharge units during the precharge operation (similarly to LISA-LIP~\cite{LISA}).
Our evaluations in Section~\ref{singlecore} show that these three changes in the timing parameters have an overall positive impact on system-level performance.



\par \noindent\textbf{Early Termination of Charge Restoration.} As described in Section~\ref{earlyprecharge}, we reduce \textbf{tRAS} and \textbf{tWR} in high-performance mode by terminating charge restoration early. Table~\ref{tab:timing_overview} shows the reduction in \textbf{tRAS} and \textbf{tWR} with by applying early termination of charge restoration (w/ E.T. column) compared to without applying it (w/o E.T. column). We find that applying early termination of charge restoration reduces \textbf{tRAS} even further and enables the reduction of \textbf{tWR}\footnote{
Compared to the conventional DRAM architecture that uses only one SA per cell, write operations to a  high-performance row in \mechanism{} require a single write driver to drive \emph{two} coupled SAs. Without terminating charge restoration early, the additional load on the write driver offsets the write latency (\textbf{tWR}) benefits of using coupled cells and SAs in high-performance mode.
}, while increasing \textbf{tRCD} by only 0.1\,ns. The marginal increase in \textbf{tRCD} is because the reduced charge level in the charged cell of the two coupled cells reduces the initial $\Delta V$ developed across the bitline pair at the very beginning of a row activation, thus extending the length of the charge sharing phase.

\subsection{Refresh Interval Versus Latency} \label{SPICE-sensitivity}

\par Section~\ref{reducerefresh} introduces the fundamental trade-off between the extended refresh interval (\textbf{tREFW}) and access latency (\textbf{tRCD} and \textbf{tRAS}). In this section, we analyze the sensitivity of \textbf{tRCD} and \textbf{tRAS} to the increase in the refresh interval.

\par
\begin{figure}[h]
\centering
\hspace*{-.1in}
\includegraphics[width=\linewidth]{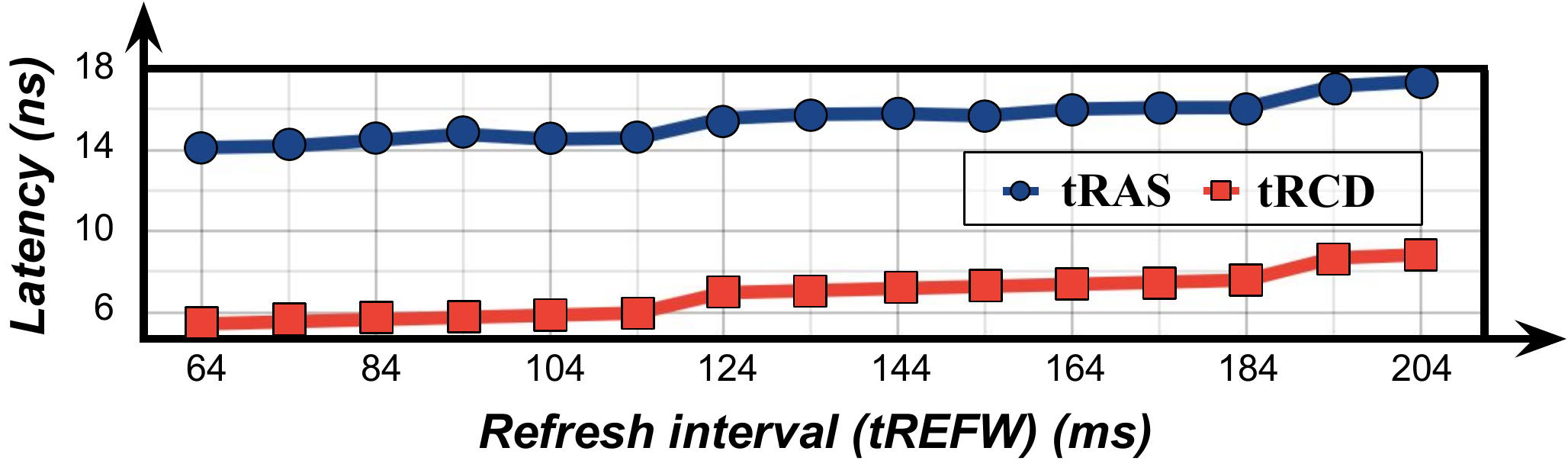}
\caption{Sensitivity of \textbf{tRCD} and \textbf{tRAS} to refresh interval.
}
\label{fig:sen}
\end{figure}

 We perform a sweep of the time interval between consecutive refresh operations to the same row (\textbf{tREFW}) and study how changing \textbf{tREFW} affects \textbf{tRCD} and \textbf{tRAS} (with early-termination of charge restoration applied). We sweep the refresh interval in increments of 10ms until the reduced charge level in the cell prior to activation  is too low for the SA to sense correctly. Figure~\ref{fig:sen} shows the results of our sensitivity analysis. We make two observations.


\par First, we observe that extending the refresh interval increases \textbf{tRCD} and \textbf{tRAS}. This is because extending the refresh interval reduces the charge level in the cell prior to activation, which extends the charge sharing phase. When the refresh interval is extended from 64ms to 194ms (a 3.03$\times$ increase), \textbf{tRCD} and \textbf{tRAS} increase by 3.24ns and 3.04ns (by 1.58$\times$ and 1.21$\times$), respectively. 

\par Second, these increases in \textbf{tRCD} and \textbf{tRAS} are large enough to potentially degrade system-level performance. 
To study the system-level effect of the trade-off between reducing the refresh rate and increasing \textbf{tRCD} and \textbf{tRAS}, we perform a sensitivity analysis using the timing parameters from Figure~\ref{fig:sen} in Section~\ref{sensystem}.


\section{System-Level Evaluation} 
\label{Evaluation}

We use system-level performance and power simulation to evaluate \mechanism{} on single- and multi-core systems.

\subsection{Methodology} \label{subsec:methodology}
\noindent \textbf{Simulators.} We use a customized version of Ramulator~\cite{kim2015ramulator}, a cycle-accurate DRAM simulator, to evaluate \mechanism{}'s system-level performance. We use Ramulator's CPU-trace driven simulation mode with application traces generated from a custom Pintool~\cite{reddi2004pin, pintool}. We use DRAMPower~\cite{chandrasekar2012drampower} to evaluate the power and energy consumption of \mechanism{} with the DRAM command traces generated by Ramulator.

\noindent \textbf{System Configuration.} The configuration of the system used in our evaluations is shown in Table~\ref{tab:Methodology}.

\label{singlecore}
\begin{figure*}[t]
\centering
\includegraphics[width=0.95\linewidth]{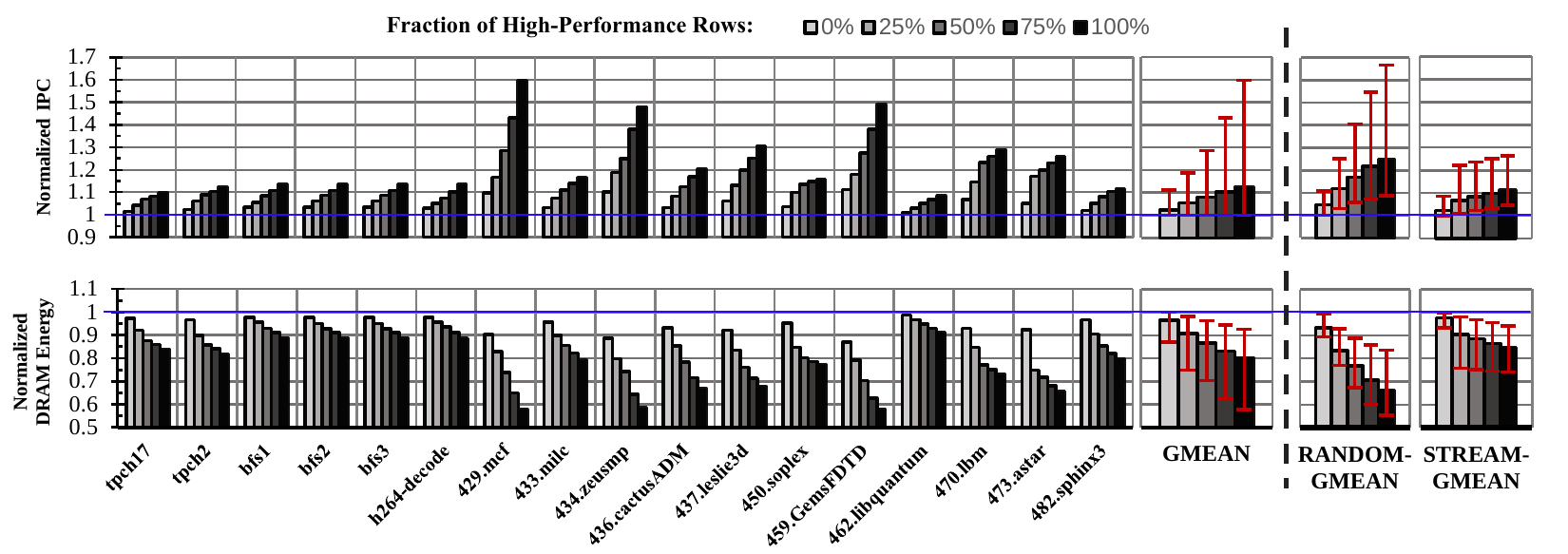} 
\caption{Normalized IPC and DRAM energy consumption of memory-intensive single-core benchmarks with \mechanism{}.}
\label{fig:individual}
\end{figure*}

\begin{table}[h]
\vspace{1mm}
\centering
\scriptsize
\setlength{\tabcolsep}{1pt}
\caption{Simulation configuration.}
\label{tab:Methodology}
\vspace{-1mm}
\begin{tabular}{ll}
\toprule
\multirow{2}{*}{\textbf{Processor}} & 1-4 core(s), 4GHz, 4-wide issue, 8 MSHRs per core \\  & 128-entry instruction window \\ \hline
\textbf{LLC} & 64-byte cacheline, 8-way associative, 8MB total capacity \\ \hline
\textbf{Memory} & FR-FCFS-Cap scheduling policy~\cite{mutlu2007stall}, timeout-based row policy\footnote{A row-buffer management policy that closes an open row after 120\,ns if there are no outstanding requests to that row.}, \\ 
\textbf{Controller} & 64-entry read/write request queue\\ \hline
\textbf{DRAM} & \begin{tabular}[c]{@{}l@{}}1 channel, 1 rank, DDR4~\cite{jedec-ddr4}, 1200MHz bus frequency, \\ 16Gb chip density,  4 bank groups, 4 banks per bank group\end{tabular} \\ \hline
\textbf{Benchmarks} &\begin{tabular}[c]{@{}l@{}} 41 benchmarks from SPEC CPU2006~\cite{spec2006}, TPC~\cite{tpc}, MediaBench~\cite{fritts2005mediabench} \\ 30 in-house synthetic random-access and stream-access traces\end{tabular} \\
\toprule
\end{tabular}%
\end{table}

\noindent \textbf{Workloads.} Our evaluation includes 1) 41 real applications from the SPEC CPU2006~\cite{spec2006}, TPC~\cite{tpc} and MediaBench~\cite{fritts2005mediabench} benchmark suites and 2) 30 in-house synthetic random and stream access workloads~\cite{Hassan2019CROW, mutlu2007memory, kim2011thread}. The random workloads randomly access memory locations, and thus exhibit a 
very limited row locality. The stream workloads access contiguous memory locations, exhibiting high
row locality.

Our single-core evaluation covers all 71 workloads. For multi-core evaluation, we first categorize the 41 real applications as memory-intensive or non-memory-intensive based on the misses-per-kilo-instruction (MPKI) metric from the last-level-cache. MPKI is calculated using the SimPoint~\cite{simpoint} traces of the representative phases of each application in single-core simulations. Applications with MPKI $>2.0$ are classified as memory-intensive. We form three multi-programmed four-core workload groups, each containing 30 workloads consisting of four randomly-selected single-core applications. The three groups are: 1) Low memory intensity (``L'') that contains four non-memory-intensive applications, 2) Medium memory intensity (``M'') that contains two non-memory-intensive and two memory-intensive applications, and 3) High memory intensity (``H'') that contains four memory-intensive applications. We simulate the multi-core workloads until each CPU core completes at least 200 million instructions. For all configurations, we initially warm up the caches by fast-forwarding 100 million instructions.

\noindent \textbf{Baseline DRAM Timing Parameters.} We derive the DRAM timing parameters related to accessing the DRAM cell array (\textbf{tRCD}, \textbf{tRAS}, \textbf{tRP} and \textbf{tWR}) from our SPICE simulations explained in Section~\ref{SPICE}. We obtain the other DRAM timing parameters from the datasheet of a commodity 16Gb DDR4 DRAM chip~\cite{Samsung-ddr4-16Gb}.

\noindent \textbf{\mechanism{} Parameters. }
We obtain \textbf{tRCD}, \textbf{tRAS}, \textbf{tRP} and \textbf{tWR} for max-capacity and high-performance rows from our SPICE simulations (Section~\ref{SPICE}). We always apply early termination of charge restoration to reduce \textbf{tRAS} and \textbf{tWR} because early termination is largely beneficial aside from a marginal increase in \textbf{tRCD}. To calculate \textbf{tRFC} for high-performance rows, we reduce the default \textbf{tRFC} by a factor equal to the average reduction in \textbf{tRAS} and \textbf{tRP}.

\noindent \textbf{\mechanism{} Data Mapping.} We model 4 different configurations representing how \mechanism{} could be configured in a system where 25\%, 50\%, 75\%, 100\% of all the DRAM rows are high-performance rows. We use a profiling-based approach (similar to prior works~\cite{Son2013CHARM, choi2015multiple}) to assign a workload's X\% of the most frequently-accessed pages to high-performance rows when the system configures X\% of all rows to operate in high-performance mode. We evaluate another configuration where \emph{all} DRAM rows operate in the max-capacity mode (we denote it as 0\%). 



\noindent \textbf{Metrics. } We use the instructions per cycle (IPC) and the weighted speedup~\cite{snavely2000symbiotic,eyerman2008system} metrics to measure the performance of single-core and multi-core systems, respectively. 
We normalize the IPC, weighted speedup and energy consumption numbers of \mechanism{} to the baseline DDR4~\cite{jedec-ddr4} DRAM to show the performance and energy impact of our design. 
All average values we present are geometric means.

\subsection{Single-core Performance} 

\par 

Figure~\ref{fig:individual} (top) shows \mechanism{}'s normalized IPC with the five different page mapping configurations (0\% to 100\%) of memory intensive benchmarks and synthetic (random and stream access) workloads relative to the baseline DRAM. Figure~\ref{fig:individual} (bottom) shows the DRAM energy consumption of the same \mechanism{} configurations, normalized to the baseline. The error bars show the maximum/minimum IPC improvement or DRAM energy on each plot. We show detailed results for the 17 benchmarks with the highest MPKI values. \textbf{GMEAN} corresponds to average results across all 41 benchmarks. 
We make five key observations from these results.

\par First, \mechanism{} improves the overall performance of our single-core workloads by 5.5\%, 7.9\%, 10.3\%, and 12.4\% when 25\%, 50\%, 75\%, and 100\% of the frequent accessed pages are mapped to high-performance mode rows, respectively. No workload experiences slowdown with \mechanism{}. \texttt{429.mcf}, with all its pages mapped to high-performance rows, achieves the highest speedup of 59.8\%. We conclude that \mechanism{} significantly improves single-core performance.

\par Second, \mechanism{} significantly reduces DRAM energy consumption. By mapping 25\%, 50\%, 75\%, and 100\% of the frequently-accessed pages to high-performance rows, \mechanism{} achieves overall DRAM energy savings of 9.2\%, 13.3\%, 16.9\%, and 19.7\%, respectively. The energy savings mainly stem from the reduced execution time of the workloads and reduction in DRAM power consumption (Section~\ref{power}).

\par Third, different workloads exhibit different sensitivities to reduced DRAM access latency. 
The results from synthetic workloads clearly show that \mechanism{} benefits workloads with random access patterns, which frequently experience row buffer conflicts, due to the significant reduction in \textit{\textbf{tRAS}} and \textit{\textbf{tRP}} that determine how quickly a row can be opened and closed.

\par Fourth, different workloads exhibit different scaling behavior as we increase the fraction of frequently-accessed pages mapped to high-performance rows. For example, some workloads (e.g., \texttt{429.mcf}, \texttt{462.libquantum}) show near-linear scaling because their memory accesses are distributed relatively evenly across their memory footprint (e.g.,  the top 25\%, 50\% and 75\% most accessed pages of \texttt{462.libquantum} cover 26.4\%, 51.2\% and 75.6\% of its memory accesses, respectively). Some others (e.g.,  \texttt{450.soplex}, \texttt{470.lbm}) scale sub-linearly 
because most of their memory accesses are concentrated in a \emph{small} fraction of their memory footprint (e.g., 85.2\% of \texttt{450.soplex}'s memory references fall into the top 25\% of its most accessed pages).
Mapping more pages to high-performance rows provides diminishing speedups in a workload that accesses memory relatively unevenly throughout its memory footprint.

\par Fifth, operating all rows in max-capacity mode has a small but clearly positive performance impact (an average of 2.4\% performance improvement) compared to the baseline DRAM for single-core workloads. This is because coupling two precharge units enables a large reduction in \textbf{tRP} despite the increased  \textbf{tRAS} and \textbf{tWR} caused by the bitline mode select transistors. 
 Average DRAM energy consumption is reduced by 3.5\%, due to the reduction in \textbf{tRP}.
We conclude that, even in the worst case when all rows are configured to operate in max-capacity mode, overall system performance and energy impact of \mechanism{} is still \emph{positive}.

\par Overall, \mechanism{} significantly improves both the performance and energy consumption of single-core workloads, with increasing benefits as more pages get mapped to high-performance rows.
\subsection{Multi-core Performance}\label{multicore}
\par
\begin{figure}[H]
\centering
\includegraphics[width=0.9\linewidth]{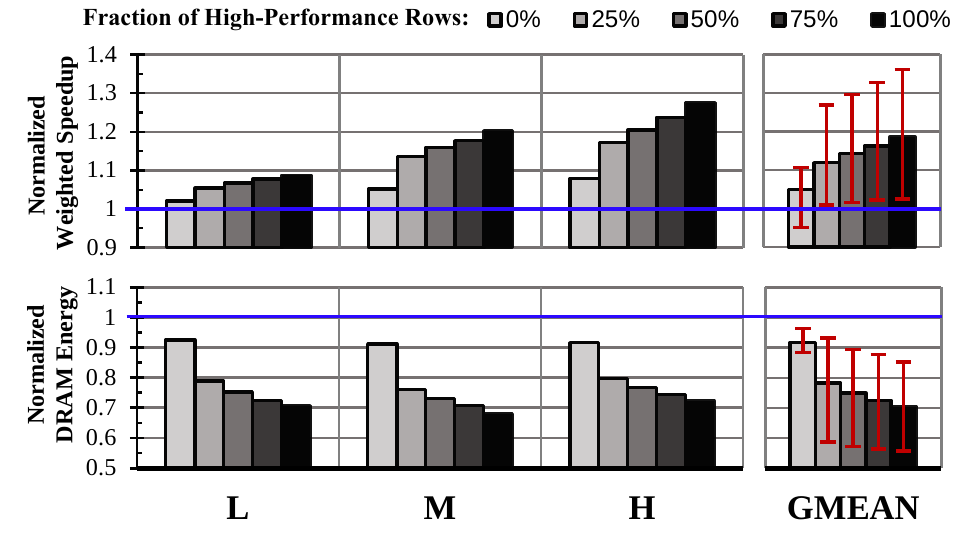} 
\caption{Normalized weighted speedup and DRAM energy consumption of \mechanism{} for 90 multi-core workloads.}
\label{fig:single}
\end{figure}

 Figure~\ref{fig:single} shows \mechanism{}'s weighted speedup (top)   and DRAM energy consumption (bottom), both normalized to  baseline DDR4 DRAM, across the three workload groups we evaluate in multi-core simulations. We make the following three observations.

\par First, \mechanism{} provides significant overall performance improvement and DRAM energy savings across all 90 multi-core workloads we evaluate. When only 25\% of the most-accessed pages are mapped to high-performance rows, \mechanism{} improves performance by 11.9\% on average
and DRAM energy consumption by 21.7\%. When this fraction is
100\%, \mechanism{} provides an overall performance improvement of 18.6\% and DRAM energy savings of 29.7\%.

\par Second, high-MPKI workloads (denoted H in Figure~\ref{fig:single}) benefit more from reducing the memory access latency in \mechanism{}'s high-performance mode. When all pages are mapped to high-performance rows, \mechanism{} provides an average speedup of 27.5\% for memory-intensive workloads since memory-intensive workloads are more sensitive to DRAM access latency reduction.

\par Third, \mechanism{} provides the smallest DRAM energy reduction for the high-MPKI workloads.
One reason for this is that \mechanism{} reduces \textbf{tRAS}  \emph{significantly}, which can cause a row to close early enough that a request accessing the same row misses in the row buffer when the request is issued shortly after closing the row. This results in more frequent row buffer misses and conflicts, which consume more energy than accesses to an open row, thereby offsetting the energy reduction benefits of the reduced execution time.

\par We conclude that \mechanism{} significantly improves performance and energy consumption of multi-core workloads.



\subsection{Power Savings of \mechanism{}}\label{power}
\par
\begin{figure}[!h]
\centering
\hspace{-2mm}
\includegraphics[width=0.9\linewidth]{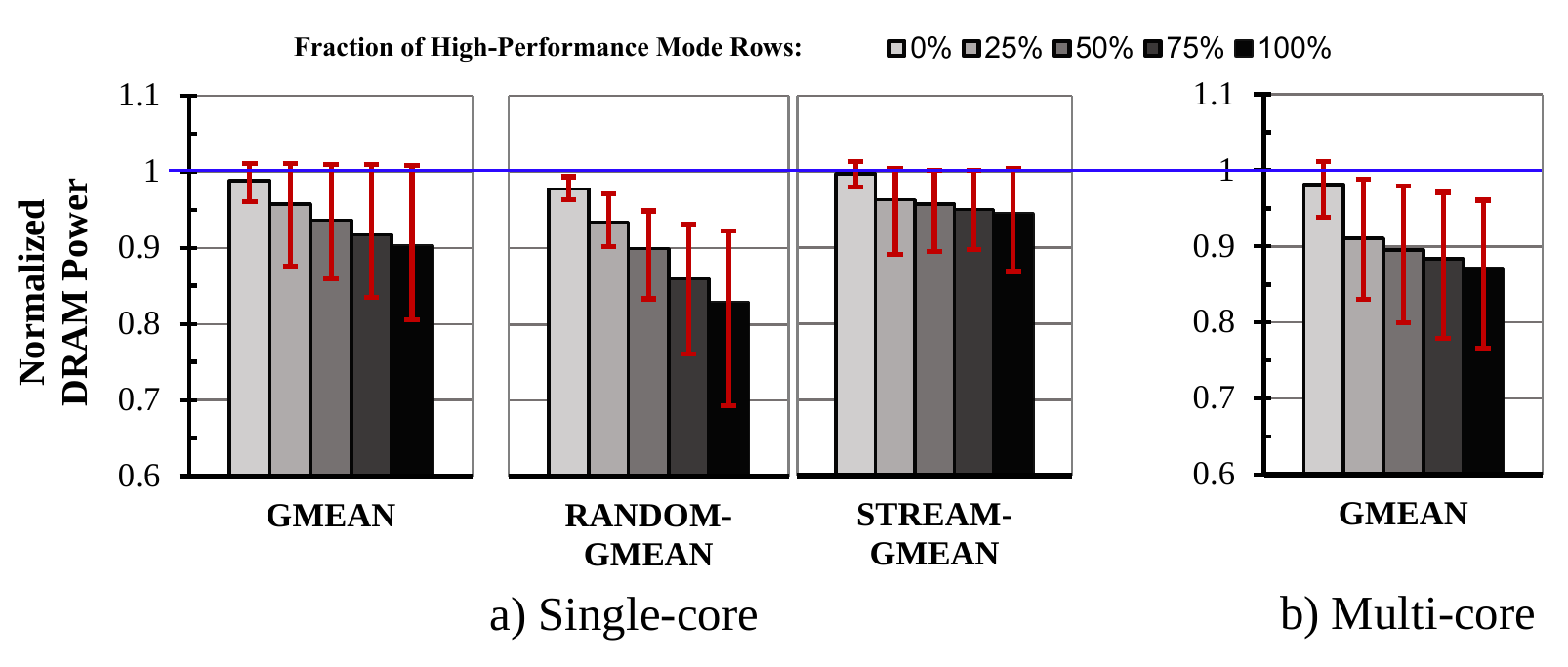} 
\vspace{-0.5em}
\caption{Normalized DRAM power of \mechanism{} for\\a) single-core and b) multi-core workloads.}
\label{fig:pwrsav}

\end{figure}

 Figures~\ref{fig:pwrsav}a and \ref{fig:pwrsav}b show the DRAM power consumption of \mechanism{} normalized to the baseline DDR4 configuration in single-core and multi-core systems, respectively. When 25\% of pages are mapped to high-performance DRAM rows, average DRAM power reduces by 4.3\% for single-core workloads (8.9\% for multi-core workloads). The average DRAM power reduction increases to 9.7\% and 12.8\% for single- and multi-core workloads, respectively, when the fraction of pages mapped to high-performance rows is 100\%. We conclude that \mechanism{} significantly reduces DRAM power consumption for both single-core and multi-core workloads. 

\subsection{Refresh Interval vs. Access Latency Trade-off}\label{sensystem}

\par Using two DRAM cells together as a logical cell with larger capacitance compared to the baseline, \mechanism{} supports extending the default refresh interval (\textbf{tREFW}) for high-performance rows. However, extending \textbf{tREFW} slightly increases \textbf{tRCD}, \textbf{tRAS}, and \textbf{tRFC} (as discussed in Section~\ref{reducerefresh}), which could potentially have a negative impact on system performance. We evaluate the performance and DRAM energy impact of increasing the refresh interval in \mechanism{}.
\par 
\noindent \textbf{Methodology.}
We evaluate four representative refresh interval \textbf{(tREFW)} settings: 114\,ms, 124\,ms, 184\,ms, and 194\,ms, each with the corresponding \textbf{tRCD} and \textbf{tRAS} values extracted using our SPICE simulations in Section~\ref{reducerefresh}. We denote these four settings as \textit{\textbf{CLR-114}}, \textit{\textbf{CLR-124}}, \textit{\textbf{CLR-184}}, \textit{\textbf{CLR-194}}. 
We perform single-core and multi-core simulations using the same methodology as described in Section~\ref{subsec:methodology} (we do not evaluate the 0\% case as max-capacity mode does \emph{not} support extending \textbf{tREFW}).
\par
\begin{figure}[h]
\centering
\includegraphics[width=\linewidth]{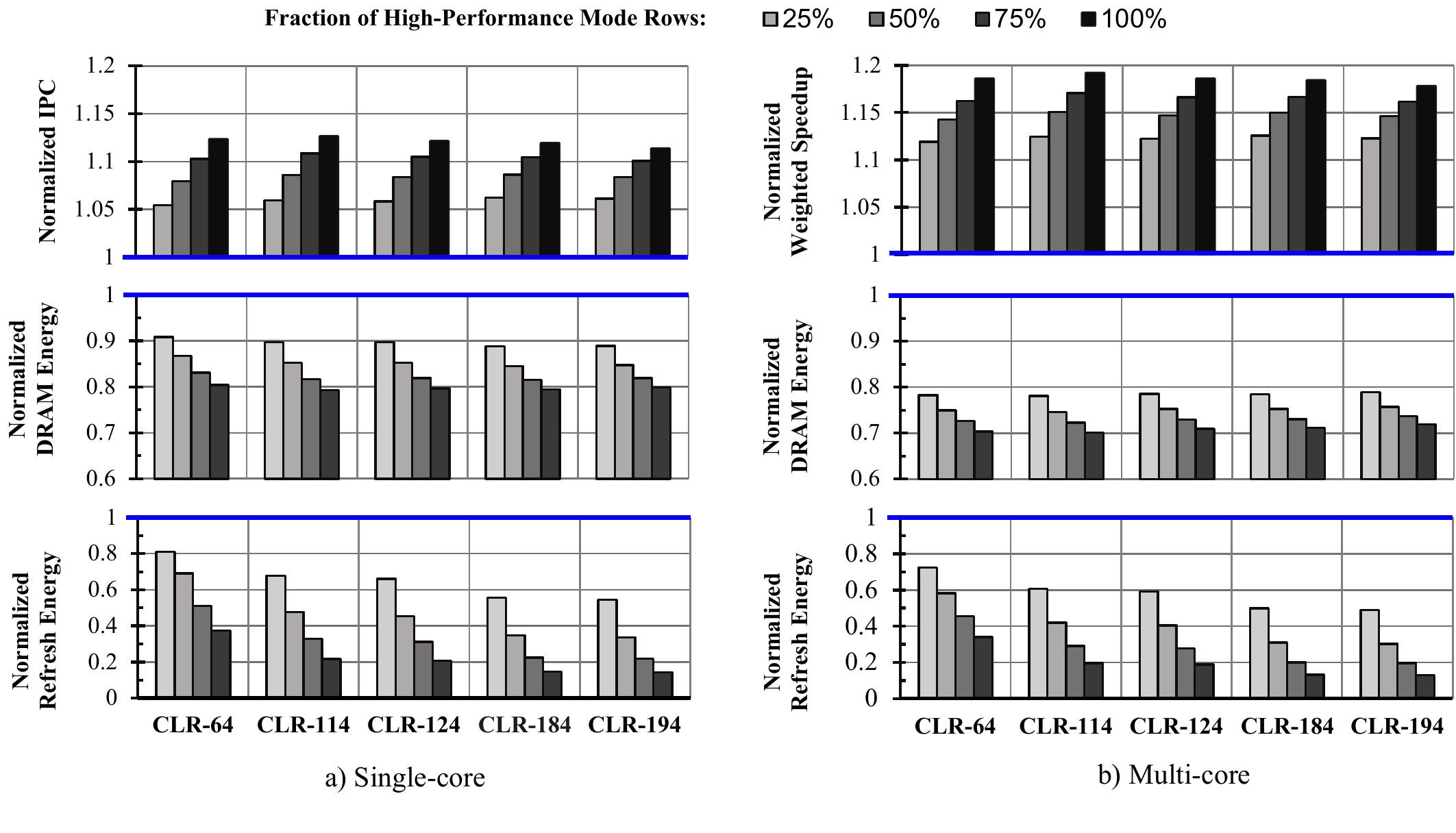} 
\vspace{-1.5em}
\caption{Normalized performance, DRAM energy, and refresh energy versus CLR-DRAM refresh interval.}

\label{fig:refe}
\end{figure}

\noindent \textbf{Results. }
Figures~\ref{fig:refe}a and \ref{fig:refe}b show the normalized IPC (weighted speedup for multi-core workloads), DRAM energy consumption, and DRAM refresh energy consumption of the four \textbf{tREFW} settings alongside the baseline case of \textit{\textbf{CLR-64}} for single- and multi-core workloads. We make the following three observations.

\par First, extending the refresh interval can either have a slightly positive or negative impact on the system performance depending on by how much the refresh interval increases. For single- and multi-core workloads, \textbf{CLR-114} provides an average speedup of 12.6\% and 19.2\% over baseline DDR4, respectively, when all pages are mapped to high-performance rows, outperforming \textbf{CLR-64} by 0.28\% and 0.61\%. In contrast, \textbf{CLR-194} causes an average slowdown of  0.98\% and 0.77\% compared to \textbf{CLR-64} (due to longer access latencies), but still outperforms the baseline by 11.4\% and 17.8\% for single- and multi-core workloads respectively.

\par Second, the overall DRAM energy consumption of all configurations is similar. This is because the energy spent on DRAM access (not DRAM refresh) is the majority of the total DRAM energy consumption. \textbf{CLR-114} provides the highest DRAM energy reduction of 29.9\% for multi-core workloads.


\par Third, by reducing \textbf{tRFC} and increasing \textbf{tREFW}, \mechanism{} saves significant DRAM refresh energy. We observe that  \textbf{CLR-64} \emph{without} an extended \textbf{tREFW} already reduces refresh energy by 66.1\% for multi-core workloads when all DRAM rows operate in high-performance mode because it uses a smaller \textit{\textbf{tRFC}}. By extending the refresh interval to 194\,ms (\textbf{CLR-194}), \mechanism{} reduces DRAM refresh energy by 87.1\%.

\par We conclude that extending the refresh interval (\textbf{tREFW}) in \mechanism{}'s high-performance mode effectively reduces the energy overhead of DRAM refresh with negligible performance impact.
\section{Related Work}


\par To our knowledge, this is the first work to provide a DRAM architecture that enables dynamic reconfiguration of any row to provide either high storage capacity or low access latency at low hardware cost. \mechanism{} provides this new dynamic capability using simple modifications to a density-optimized commodity DRAM chip.
We briefly discuss related works that exploit static and dynamic DRAM capacity-latency trade-offs, enable in-DRAM caching, reduce access latency, and mitigate refresh costs.



\noindent
\textbf{Static Capacity-latency Trade-off.} 
Certain commercial DRAM chips~\cite{rldram, lldram, hpdaram} reduce access latency by incorporating small DRAM subarrays and banks. These products \emph{statically} exploit the capacity-latency trade-off by sacrificing significant storage capacity for reduced latency. CHARM~\cite{Son2013CHARM} and TL-DRAM~\cite{lee2013tiered} \emph{statically} partition a DRAM chip into a fast region and a slow region. However, the low-latency regions in these designs are \emph{small} and \emph{fixed}.
 
\par
Designs based on Twin-Cell DRAM~\cite{takemura20060, twincell, Kim2006TwinCell} \emph{statically} couple every two DRAM cells to reduce access latency and refresh rates. However, Twin-Cell DRAM designs do \emph{not} couple two sense amplifiers to accelerate the activation and precharge process, which significantly limits their potential to improve DRAM latency compared to \mechanism{}. 
 
\noindent
\textbf{Dynamic Capacity-latency Trade-offs.}  MCR DRAM~\cite{choi2015multiple} simultaneously activates multiple (i.e., two or four) DRAM rows to achieve low row activation latency. This enables a dynamic trade-off between DRAM capacity and performance comparable to our design. However, MCR DRAM does not couple two sets of sense amplifiers and precharge units, which greatly limits the amount of latency reduction it can achieve.

\par Hsu et al.~\cite{Hsu2001intertwin} patent a DRAM array design that provides rows that are interchangeable between single-cell and twin-cell operation. Unfortunately, their design adds a new row of SAs per subarray \emph{dedicated} to twin-cell operation, which results in high DRAM chip area overhead. Their design also does not couple two SAs, so it has limited potential to reduce DRAM access latency compared to \mechanism{}.



\noindent
\textbf{In-DRAM Caching.} SALP~\cite{kim-isca2012} utilizes the parallelism between subarrays to enable simultaneously opening multiple DRAM rows in different \emph{independent} subarrays. CROW~\cite{Hassan2019CROW} enables in-DRAM caching by copying the data of a frequently-accessed DRAM row to another row (i.e., \emph{copy-row}) in the same subarray. CROW simultaneously activates the two rows that contain the same data to reduce DRAM access latency. To reduce DRAM refresh overhead, CROW eliminates weak rows, which have short retention times, and extends the refresh interval by remapping the weak rows to copy-rows that can retain data for a long time. LISA~\cite{LISA} enables fast data movement across different subarrays in a bank at the granularity of DRAM rows. The LISA-VILLA mechanism utilizes this fast in-bank data movement to cache frequently-accessed data using a small but low-latency subarray within the same bank. Cache DRAM~\cite{hidaka1990cache} adds a small SRAM cache in a DRAM chip. Because \mechanism{} enables a new row-granularity capacity-latency trade-off, \mechanism{} is orthogonal to in-DRAM caching, and mechanisms such as SALP, CROW, LISA, and Cache DRAM can be built on top of the \mechanism{} architecture to improve performance even further.


\noindent
\textbf{Reducing Latency.} Various works propose techniques to reliably reduce DRAM latency without sacrificing capacity~\cite{AL-DRAM, zhang2016restore, hassan2016chargecache, FLYDRAM, lee2016reducing, kim2018solar, wang2018reducing, chang2017understanding} by reducing timing margins under certain conditions. \mechanism{} is complementary to mechanisms proposed by these prior works and can be combined with them to reap the benefits of both. 

\noindent
\textbf{Mitigating Refresh Costs.} Several prior works provide mechanisms for reducing the performance and energy consumption penalties incurred by DRAM refresh~\cite{venkatesan2006retention, liu2012raidr, nair2013archshield, qureshi-dsn2015, ohsawa1998optimizing, wang2014proactivedram,lin2012secret, patel2017reach, khan-sigmetrics2014, das2018vrldram, chang2014improving}. Unlike \mechanism{}, these proposals do not reduce DRAM access latency.
\section{Conclusion}

We propose \mechanism{}, new DRAM architecture that enables dynamic fine-grained reconfigurability between high-capacity and low-latency operation. \mechanism{} can reconfigure every \emph{single DRAM row} to operate in either max-capacity mode or  high-performance mode. A max-capacity row maintains approximately the same storage density as the commodity density-optimized open-bitline architecture by letting each DRAM   cell operate separately. A high-performance row provides low access latency and low refresh overhead by coupling every two adjacent DRAM cells in the row and their sense amplifiers.
Our evaluations show that \mechanism{}  significantly improves system performance  and energy consumption for both single- and multi-core workloads in all our evaluation configurations, including those that favor maximizing capacity and minimizing latency.
We hope that future work exploits \mechanism{} to develop more flexible systems that can adapt to the diverse and dynamically changing DRAM capacity and latency demands of workloads.




\section*{Acknowledgments} We thank the anonymous ISCA 2020 reviewers for their
feedback and the SAFARI group members for the stimulating intellectual environment they provide. We acknowledge the generous gifts provided by our industrial partners: Alibaba, Facebook, Google, Huawei, Intel, Microsoft, and VMware.



%



\SetTracking
 [ no ligatures = {f},
 outer kerning = {*,*} ]
 { encoding = * }
 { -40 } 

{
  \let\OLDthebibliography\thebibliography
  \renewcommand\thebibliography[1]{
    \OLDthebibliography{#1}
    \setlength{\parskip}{0pt}
    \setlength{\itemsep}{0pt}
  }
  \bibliographystyle{IEEEtranS.bst}
\bibliography{ref}

}

\end{document}